%% file: main.tex
\documentclass[sigconf]{acmart}

\newif\ifnotes
\notestrue

\graphicspath{{.}{./}{./figs/}}

\usepackage{subcaption}
\usepackage{color,soul}
\usepackage{algorithm}
\usepackage[noend]{algpseudocode}
\usepackage{wrapfig,lipsum,booktabs}
\usepackage[utf8]{inputenc}

\usepackage{amsmath}

\AtBeginDocument{%
  \providecommand\BibTeX{{%
    \normalfont B\kern-0.5em{\scshape i\kern-0.25em b}\kern-0.8em\TeX}}}

\setcopyright{acmcopyright}
\copyrightyear{2021}
\acmYear{2021}
\acmDOI{10.1145/1122445.1122456}

\acmConference[CHI '21]{CHI '21: ACM Conference on Human Factors in Computing Systems}{May 08--13, 2021}{Yokohama, Japan}
\acmBooktitle{CHI '21: ACM Conference on Human Factors in Computing Systems,
  May 08--13, 2018, Yokohama, Japan}
\acmPrice{15.00}
\acmISBN{978-1-4503-XXXX-X/18/06}

\acmSubmissionID{3295}

\begin{document}

%%
%% The "title" command has an optional parameter,
%% allowing the author to define a "short title" to be used in page headers.
\title{\textit{User Ex Machina} : Simulation as a Design Probe in Human-in-the-Loop Text Analytics}

%%
%% The "author" command and its associated commands are used to define
%% the authors and their affiliations.
%% Of note is the shared affiliation of the first two authors, and the
%% "authornote" and "authornotemark" commands
%% used to denote shared contribution to the research.

\author{Anamaria Crisan}
\affiliation{%
  \institution{Tableau Research}
  \city{Seattle}
  \country{USA}
}
\email{acrisan@tableau.com}

\author{Michael Correll}
\affiliation{%
  \institution{Tableau Research}
  \city{Seattle}
  \country{USA}
}
\email{mcorrell@tableau.com}

%%
%% By default, the full list of authors will be used in the page
%% headers. Often, this list is too long, and will overlap
%% other information printed in the page headers. This command allows
%% the author to define a more concise list
%% of authors' names for this purpose.
\renewcommand{\shortauthors}{Crisan and Correll, et al.}

%%
%% The abstract is a short summary of the work to be presented in the
%% article.
\input{text/0-abstract}
%%
%% The code below is generated by the tool at http://dl.acm.org/ccs.cfm.
%% Please copy and paste the code instead of the example below.
%%


%%
%% Keywords. The author(s) should pick words that accurately describe
%% the work being presented. Separate the keywords with commas.

%% A "teaser" image appears between the author and affiliation
%% information and the body of the document, and typically spans the
%% page.
\begin{teaserfigure}
\centering
\includegraphics[width=\textwidth]{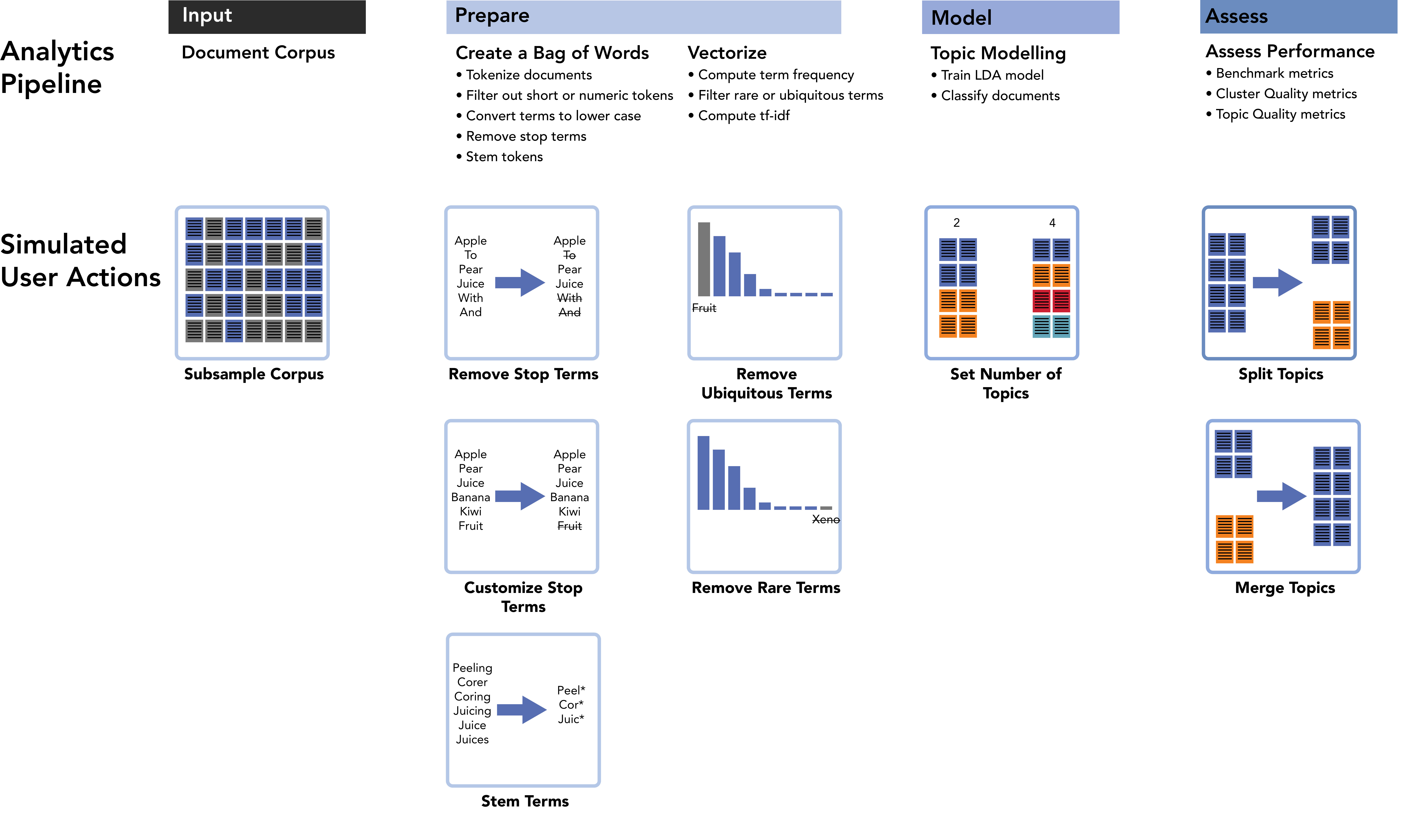}
%    \begin{subfigure}[c]{0.45\textwidth}
%    \includegraphics[width=\textwidth]{LDA_pipeline}
%    \end{subfigure}
%    ~
%    \begin{subfigure}[c]{0.55\textwidth}
%    \includegraphics[width=\textwidth]{actions}
%    \end{subfigure}
  \caption{On the top, our topic modelling pipeline. On the bottom, user actions that we simulate aligned to the respective pipeline phase. Some of these actions occur before the first model is generated; others might occur during or after the modelling stage, in order to improve or adjust the results. We should \textit{quantify} and \textit{clearly communicate} the impact of these sorts of actions on the resulting model.\\
  %\mc{If we are not splitting modelling and assessing, then I need to split up the bottom row here to match fig \protect\ref{fig:lda_pipeline}}
  }
  \Description{The eight user actions modeled in our pipeline, divided by whether they are actions that occur in the input, preparation, modelling, or assessing stages. These actions are described non-graphically later in section 3 of the paper.}
  \label{fig:teaser}
\end{teaserfigure}

%%
%% This command processes the author and affiliation and title
%% information and builds the first part of the formatted document.
\maketitle

\input{text/1-intro}
\input{text/2-related}

\input{text/3-methods}
\input{text/4-results}
\input{text/5-discussion}

%%
%% The acknowledgments section is defined using the "acks" environment
%% (and NOT an unnumbered section). This ensures the proper
%% identification of the section in the article metadata, and the
%% consistent spelling of the heading.
\begin{acks}
We would like to acknowledge our colleagues at Tableau and elsewhere who provided early feedback and motivating examples for this work: Zo\"{e}, Eric Alexander, Michael Arvold, Dan Cory, Kate Mann, Zach Morrissey, Britta Nielsen, Vidya Setlur, and Maureen Stone. We also thank the reviewers for their feedback.  
\end{acks}

%%
%% The next two lines define the bibliography style to be used, and
%% the bibliography file.
\bibliographystyle{ACM-Reference-Format}
\bibliography{main}

%%
%% If your work has an appendix, this is the place to put it.
%\appendix

\end{document}

%% file: text/0-abstract.tex
\begin{abstract}
  Topic models are widely used analysis techniques for clustering documents and surfacing thematic elements of text corpora. These models remain challenging to optimize and often require a ``human-in-the-loop'' approach where domain experts use their knowledge to steer and adjust. However, the fragility, incompleteness, and opacity of these models means even minor changes could induce large and potentially undesirable changes in resulting model. In this paper we conduct a simulation-based analysis of human-centered interactions with topic models, with the objective of measuring the sensitivity of topic models to common classes of user actions. We find that user interactions have impacts that differ in magnitude but often negatively affect the quality of the resulting modelling in a way that can be difficult for the user to evaluate. We suggest the incorporation of sensitivity and ``multiverse'' analyses to topic model interfaces to surface and overcome these deficiencies.
  \\
  %I got dinged by lazy reviewers that claimed I didn't have supplemental materials.
  \textbf{Code and Data Availability}: \texttt{\url{https://osf.io/zgqaw}}
\end{abstract}

%% file: text/1-intro.tex
\section{Introduction}
%\ac{Modified the way results were written here, removed some redundancy in the text.}

Entire domains of scholarship are dedicated to the semantic analysis of text. Attempts to support and augment these processes of human interpretation and summarization computationally have often struggled with the degree to which human agency should shape or control the algorithmic output. On one extreme there are positions such as the quote often ascribed to Frederick Jelinek~\cite{hirschberg1998every} that ``every time I fire a linguist, the performance of the speech recognizer goes up,'' suggesting that the role of human expertise may be somewhat limited, and perhaps even counter-productive in the face of algorithmic complexity and performance beyond the capacity of the usual interpreters of texts. On the other hand, the opaque, biased, and brittle approaches of computational and statistical models~\cite{o2016weapons} has led to calls for explainable AI (XAI) and human-in-loop machine learning (HILML). 

Text corpora are often of such size and complexity that we cannot read or analyze all the texts therein. Computational ``distant reading''~\cite{moretti2005graphs} approaches such as topic modelling can allow us to form an impression of the content or important patterns in text corpora without reading every document. Human agency in building, interpreting, and communicating the results of these text models is an important component of their use~\cite{correll2012shakespeare}. The specific role of this human agency can take many forms: Lee et al.~\cite{lee2019human} propose three potential levels of autonomy in model-building, from entirely user-driven exploration of the model space, to a ``cruise control'' level where the user provides periodic coarse guidance while the system adapts the fine-grain details, all the way up to full ``autopilot'' where the system has full control over the model output. Likewise, Heer~\cite{heer2019agency} suggests a hybrid approach where automated methods ``augment'' (but do not remove agency) from human analytical decisions. However, despite human agency in their creation, resulting models for text analytics may be brittle, difficult to interpret, or fail to capture semantic information of relevance to the reader. How these text analytics models are \textit{built} versus how they are \textit{interpreted} may be at odds~\cite{chang_reading_2009}, resulting in ``folk theories'' of algorithmic performance that may or may not not reflect realities of algorithmic performance or structure~\cite{devito2018algorithm}.

\iffalse
\begin{figure*}
  \centering
    \begin{subfigure}[c]{0.68\textwidth}
        \includegraphics[width=\textwidth]{clusterfail}
        \caption{Altering topics}
    \end{subfigure}
    
    \begin{subfigure}[c]{0.7\textwidth}
     \includegraphics[width=\textwidth]{topicfail}
     \caption{Altering words}
    \end{subfigure}

\caption{Potential outcomes of user interactions with topic models, that were ``successful'' or ``unsuccessful'' with respect to steering the model in a human-centered way. In the first interaction, the user believes that the cluster of documents belonging to Topic 1 represent too broad of a semantic category, and so increases the number of topics. Depending on the model, this could either ``split'' Topic 1 as desired, or, in an unsupervised way, result in a totally different landscape of topic clusters, with no little relation to the initial state. Similarly, in the second interaction, the user may observe that the generated topics in a corpus of recipes are ``polluted'' by stop words, and seek to remove them for the model. If done successfully, the resulting topics are ``cleaner'' and include more semantically important words. However, removing stop words can be disruptive\protect\cite{chuang_topiccheck_2015}, and it's possible that this action would result in new topics that have no resemblance to the old.}
\label{fig:exampleActions}
\end{figure*}
\fi

We view the combination of the purported utility of human guidance in text analytics, the fragility and instability of existing corpus linguistic tools, and the potential for resulting models to be misinterpreted, as a provocation to question how \textit{disruptive} human actions may be to text models. As an example of this issue, a user could perform an action that is, from their perspective, a minor adjustment or correction based on domain knowledge that could entirely reconfigure the topic space. In such a case, it is difficult for the user to remain oriented when performing future analytic tasks with the corpus, or be able to judge whether or not their action had a positive impact on the model. A measure of user impact could be in terms familiar to evaluators of statistical models (such as complexity, accuracy, or precision), or it could be disruption in more human terms (the semantic content of particular topics, or the visual or textual summaries of those topics). There are likewise failures in the other direction: a user might perform an action that they believe will ``fix'' a consistent problem with the topic model output, but that results in irrelevant or superficial changes. Without a fuller understanding of the impact of user actions from a both algorithmic and human-centric standpoint, we risk producing steerable or supervised systems that are frustrating to use and interpret, and may not even produce the hoped-for gains in accuracy, agency, and interpretability.

In this paper, we investigate, through simulations, how efforts to steer or update topic models impact the resulting ``truthiness,'' coherence, and interpretability of the altered model. Our agenda is to determine to what degree topic models as they are commonly employed for content analysis can meaningfully respond to human actions, and the degree to which the resulting changes are robust and reliable from a human-centered standpoint: in short, we wish to perform a \textbf{human-centered sensitivity analysis of topic modelling}. We focus on potential user actions throughout the text analytics pipeline, from the way that words are prepped and modelled, to specific interactions with the model output (see \autoref{fig:teaser}). The contributions of this work are:
%\begin{itemize}
%\item A text-analytics pipeline capable of incorporating user %actions at various stages from data preparation, modelling, and %performance assessment
%\item The use of benchmark, cluster, and topic quality metrics to %assess the impact of user actions
%\item AVD representation for framing the user's expected impact of %an action against the model impact
%\end{itemize}

%MC Contributions attempt
\begin{itemize}
    \item A text analytics pipeline capable of simulating potential user actions at various stages of topic modelling, including data preparation, modelling, and performance assessment.
    \item A holistic assessment of the impact of these actions, in terms of established benchmarks like topic and cluster quality metrics as well as in terms of impact on the resulting summary visualizations.
    \item Recommendations for future designers of human-in-the-loop text analytics systems.
\end{itemize}

We deploy our simulation approach using datasets with known and absent ground-truth labels to measure the impact of user actions on model performance. We found that user actions have a range of impact, some modifying model performance very little and others causing more substantive changes. We also found that user actions that impact data preparation resulted in the largest changes in model performance, although these changes may or may not be reflected in resulting visualizations of the model.
%Moreover, we show that data visualization can potentially mislead users as to the impact of their actions, at times making the action appear more impactful than it actually is.  
Our results demonstrate the importance of giving the user agency to introduce changes across the text analytics pipeline and for designers of human-in-the-loop text analytics systems to communicate the impact of these actions through data visualization. We call on designers to: \textbf{surface} the provenance and data flow choices of inputs to topic models (not just visualizations of the \textit{output} of such models), \textbf{alert} users to potentially disruptive impacts of their decisions, and to \textbf{guide} through the simultaneous exploration of multiple analytic ``paths.'' 

%% file: text/2-related.tex
\section{Related Work}

Many text analytics systems are designed to support human agency, whether this agency takes the form of human-driven ``interaction'' with ~\cite{brown_dis-function_2012} or ``supervision'' of~\cite{mao_sshlda_2012} text models.  For instance, the iVisClassifier~\cite{choo2010ivisclassifier} system allows users to supervise the clustering of text corpora, while tools like TextTonic~\cite{paul_textonic_2019} and Dis-Function~\cite{brown_dis-function_2012} afford user-driven fine-grained steering of a clustering and layout by, for instance, ``pinning'' important words, or dragging points the user believes to be misplaced to better locations. However, there are many degrees of freedom in how text models are built, from preparation to analysis to visualization. All of these decisions could potentially benefit from human intervention: Smith et al.~\cite{smith2018} note that, though there are costs of human interaction with topic models such as latency and unpredictability, these costs can be qualitatively offset by increased perceived ownership, trust, or performance.

We focus specifically on the case of topic modelling for text analytics as a candidate for human interactions with the algorithm, and visual explanations of the resulting model. % In this work, we focus on a representative scenario of content analysis across a corpus.
We assume that the analyst is interested in viewing an \textbf{overview} of a corpus, composed of \textbf{clusters} of \textbf{texts}. We assume that membership in these clusters is driven by \textbf{topics} generated from a \textbf{topic model}. Our objective was to determine how potential human interactions might shape the resulting topic clusters in terms of accuracy, interpretability, and resiliency. Our results build upon prior considerations of topic modelling as a tool for content analysis, potential user interactions with text models, and sensitivity analyses of visual analytics designs and algorithms.

\subsection{Topic Modelling}

A common task in text analytics is determining the themes of a large text corpus: what are the texts in a corpus generally \textit{about}, and which texts are \textit{about} which topics? Beyond functioning as a way of analyzing the content of a corpus \textit{per se}, a topic model can be useful for searching for particular documents, orienting oneself in an unfamiliar text dataset, or performing data cleaning tasks (such as filtering out irrelevant or mislabeled documents). Latent dirichlet allocation (LDA)~\cite{blei2003latent} is a common statistical approach to topic modelling. The corpus is assumed to be made up of a predetermined number of \textit{topics}. Each \textit{topic} is a probability distribution across all of the \textit{tokens} (usually words) in the corpus. Texts (reduced to a ``bag-of-words'' vectorization) are then taken to be distributions across topics, as though one were drawing words out of a weighted sample of different topic boxes, each of which contains its own collection of words. By analyzing the words that are prominent in topics, and analyzing the topics that are prominent in texts, the analyst can get a picture of the content and composition of a corpus.

\subsubsection{Topic Model Visualization}
\label{sec:related-topic-model-visualization}

A full scope of text visualization techniques, even just those subset of text visualization techniques meant for content analysis, is outside of the scope of this paper (consult Kucher et al.~\cite{kucher2015text} for a survey). We instead focus on visualization techniques directly related to topic models, or the challenge of visualizing different clusters of semantic content in text corpora. We note here that our survey of methods is biased towards those where the topic models are both a tool for structuring the text corpora but also \textit{objects of inquiry themselves}. Many visualizations may not expose the inner structure of the topics at all, leaving it as a black box that is used to determine cluster membership or pairwise document distances. We identify two common clusters of designs of topic model visualizations:
%(a third cluster, visualizations of topic models \textit{over time} such as Theme River~\cite{havre2002themeriver}, we exclude for reasons of relevance): \ac{People can look at the survey}

\textbf{Topic Matrices:} assessing the utility of methods like LDA often involves examining the contribution of each word to a topic, or each text to a topic, or some other pairwise comparison of values. Tools like Termite~\cite{chuang_termite_2012} and Serendip~\cite{alexander_serendip_2014} present this information in the form of matrices of topic information. Since there may be many topics, words, and texts under consideration, a key design challenge is how to make the resulting matrix usable and interpretable by humans. Saliency metrics are often used to drive ordering or filtering of these matrices, combined with operations like roll-up and drill-down. The assumption is that the viewer may only be able to see a small fraction (say, the distribution of the top $10$ tokens across the top $10$ topics) of the matrix at once.

\textbf{2D Spatializations:} other visualization tools use topic models or other measures of text distance to power a resulting ``spatialization''~\cite{tory2007spatialization,wise1995visualizing} or ``landscape''~\cite{chalmers1993using} of the corpus. Adjutant~\cite{crisan_adjutant_2019} and the Stanford Dissertation Browser~\cite{chuang_interpretation_2012} both present the user with a two-dimensional projection of the corpus, with explicitly identified clusters that are meant to represent topics of interest. While 2D planes~\cite{crisan_adjutant_2019} and graphs~\cite{cao2010facetatlas} are standard spatializations, radial or polar views of text are also common. DocuBurst~\cite{collins2009docuburst}, TopicPie~\cite{yang_topicpie_2016}, PhenoLines~\cite{glueck_phenolines_2018} and VISTopic~\cite{yang_vistopic_2017} all employ radial views of texts or corpora. These radial views are often based around a hierarchical spatialization or organization of texts, with ``core'' topics or tokens afforded greater size or centrality than peripheral or finer grained topics.  A particular challenge with spatializations is that the space itself can have important semantic or analytic connotations~\cite{liu2019latent}: for instance, the location of one of Shakespeare's plays in a scatterplot can be interpreted as encoding information about genre~\cite{hope2010hundredth}.

%\textbf{Theme Rivers:}~\cite{havre2002themeriver} a common content analysis task is examining the changes in topics over time. Stacked area or bar charts showing the contribution of text ``facets''~\cite{shi2010understanding} over time often resemble meandering ``rivers'' as topics of interest decrease or increase in frequency over time. Other designs for visualizing facets over time, such as parallel tag clouds~\cite{collins2009parallel}, rely on juxtaposed list of words with flows between them to act as proxies for explicitly delineated topics.

%\ac{Instead of having this be a list, simplified it a bit and made it paragraph}.
These two design categories simplify information by reducing the vast amounts of data produced at the term, text, and corpus level into something more manageable for humans to review. These visualizations  focus on just a subset of the data in topic models, or rely on multiple coordinated views~\cite{alexander_serendip_2014,el-assady_progressive_2018} to present additional facets or levels of detail. However, visualizations are also sensitive to text analytics algorithms and their parameter configuration. For example, the designers of Termite~\cite{chuang_termite_2012} found that the utility of their visualization was highly dependent on the metrics they employed to order words. Parallel Tag Clouds~\cite{collins2009parallel} likewise concern themselves with how words should be ordered in their texts. Choices of dimensionality reduction~\cite{chuang_interpretation_2012} and automatic topic labeling algorithms~\cite{chuang_without_2012} can likewise impact how topics in a corpora are interpreted by the viewer. In our analysis, we explore how simulated user actions impact the two categories of text visualizations.
%There are commonalities across two design categories:
%\begin{itemize}
%    \item \textbf{Simplification}: topic models produce data at the token, text, and corpus level. Visualizing the data at each level simultaneously is often infeasible or unnecessary for particular algorithmic tasks. Visualizations often focus on just a subset of the data in topic models, or rely on multiple coordinated views~\cite{alexander_serendip_2014,el-assady_progressive_2018} to present additional facets or levels of detail.
%    \item \textbf{Sensitivity to Algorithms and Parameters:} The designers of Termite~\cite{chuang_termite_2012} found that the utility of their visualization was highly dependent on the metrics they employed to order words. Parallel Tag Clouds~\cite{collins2009parallel} likewise concern themselves with how words should be ordered in their texts. Choices of dimensionality reduction~\cite{chuang_interpretation_2012} and automatic topic labeling algorithms~\cite{chuang_without_2012} can likewise impact how topics in a corpora are interpreted by the viewer.
%\end{itemize}

%These commonalities, in concert, suggest the need to \textit{compare} topic models across different parameter setting (either analytically or visually). 
%and a potential \textit{danger} if the (simplified) visualization of a particular topic model does not communicate important features of the model, such as reliability or uncertainty.
%We explore both categories of visualizations in our analysis.

\subsubsection{Topic Model Comparison}

LDA and many other topic modelling algorithms are probabilistic and there are stochastic elements to their output~\cite{steyvers2007probabilistic}; even on the same corpus, a topic model may differ from run to run, producing substantially different or even contradictory analyses of the same corpus~\cite{roberts2016navigating}. Even without this concern, there are many degrees of freedom (the selection of the number of topics, the pre-processing of texts, the creation of the bag-of-words model) that can result in differing outputs (both in terms of the model itself, and the visualization of the model). As such, there is an interest in visually comparing two or more topic models. Alexander \& Gleicher~\cite{alexander_task-driven_2016} treat the topic model comparison task as motivation for a design exercise, creating matrix based views as well as ``buddy plots'' that allow the viewer to see how individual texts shift in topic space across models. Our interest in this space is more specific, however; we are concerned with comparing many topic models simultaneously, investigating the sensitivity of different parameter settings on these models, and comparing these models to an existing ground truth when available. As such, we selected three designs as inspiration:

TopicCheck~\cite{chuang_topiccheck_2015} uses a matrix of small multiples to assess the stability of a topic model algorithm across runs. Columns are different runs of the model, and rows are different ``groups'' of highly similar topics. By observing ``gaps'' in the matrix (where particular topics did not persist across runs), the viewer can gain some sense of the stability of a particular algorithm on the given corpus.

Resonant with our research questions, El-Assady et al.~\cite{el-assady_progressive_2018} employ a per-parameter comparison of topic models, allowing the user to gauge the potential impact of different weightings on the resulting topics. Their use case, where the analyst interactively explores the parameter space and iteratively refines the output model, closely matches our vision of a ``steerable'' topic model system.

Lastly, Chuang et al.~\cite{chuang2013topic} employ a matrix visualization with marginal bar charts to compare the results of a topic model with ``latent'' concepts (what would be the ``ground truth topics'' in our scenario). Of particular interest in their design are latent concepts that are ``missing'' (not covered by any of the generated topics), ``repeated'' (covered by multiple topics), and likewise generated topics that are ``fused'' (containing multiple latent concepts) or ``junk'' (not corresponding to any of the latent concepts). 

As with the visualization techniques for individual topic models, the comparison of two or more topic models is also highly sensitive to the choice of specific metrics employed.

\subsubsection{Topic Model Metrics}~\label{sec:related-topic-model-metrics}

Many diagnostic measures for topic models have been proposed, often relying on the probabilistic or information theoretic properties of the topics themselves. Topics, as vectors in a high-dimensional token-space, can be compared via standard vector difference measures. Beyond euclidean distance, cosine similarity~\cite{jacobi2016quantitative}, Jensen-Shannon similarity~\cite{fothergill2016evaluating}, and KL-divergence~\cite{chuang2013topic} have all been used to measure distances between topics. These metrics are employed to quantify more abstract concepts such as the \textit{coherence} of topics, the \textit{distance} between topics, or the relation of these topics to \textit{ground truth} latent concepts. As topics and bag-of-words texts are both vectors of tokens with associated weights, these metrics can also be used to measure the coherency of topics: e.g., El-Assady et al.~\cite{el-assady_progressive_2018} use a Ranked Weighted Penalty Function to both measure the distance between two topics, and the coherency of the texts within that topic. 

Within topics, there is a challenge in measuring the centrality or \textit{saliency} or particular tokens. These saliency metrics are often used to \textit{order} words within a topic (so that a final visual or textual summary can include the top $n$ most important tokens, rather than the unwieldy full list of all tokens with non-zero value) or to \textit{label} particular topics with descriptive phrases. Chuang et al.~\cite{chuang_without_2012} find that na\"ive orderings based on, e.g., term frequency may not surface the tokens in a topic that are the most effective summaries of the topic's contents. They also find that more complex metrics such as the $G^2$ measure used by other text visualization systems~\cite{collins2009parallel} may not adequately capture how humans summarize texts. The choice of word-ranking metric can have large impacts on the resulting visualization~\cite{chuang_termite_2012}, and therefore the resulting analyses based on those visualizations.

There are also human-centric metrics for assessing the coherence and utility of topics. Fang et al.~\cite{fang2016using} employ word embedding models to measure cluster coherence via the semantic similarity of its top words. Chang et al.\cite{chang_reading_2009} propose the ``word intrusion'' and ``topic instrusion'' tasks as human-derived metrics for topic models, corresponding to the reliability and ease at which humans can detect extraneous words or topics from a list. Recognizing the cost of having to gather data from human subjects, Lau et al.~\cite{lau_machine_2014} attempt to construct algorithmic measures that emulate human performance at these word intrusion tasks. Most relevant to our work, Kumar et al.~\cite{kumar2019didn} compute a ``control'' metric based on elicited or simulated priors about document ranks to measure the impact of different simulated actions on topic model outputs.

These differing approaches to human-interpretable topic metrics may often disagree with each other, be expensive to compute, or require human input or independently trained models to be feasible. We limit our simulated analyses to more standard vector distance metrics for reasons of computational efficiency, but we acknowledge that one metric is unlikely to suffice to capture the full picture of how a particular topic is perceived. 

\subsubsection{Topic Cluster Metrics}

%\mc{I am dividing this up into two classes to address your comments and to more closely mirror the methods section. Right now these are ``benchmark'' metrics (vs. ground truth) and ``cluster-to-cluster'' metrics (no ground truth). Not wild about those terms so happy to hear alternatives. Will try to make sure these changes percolate down to the rest of the doc, e.g. methods and results.}

In our specific use case, a topic model is used to generate a membership function for document clusters. As such, an analyst might be interested in the the quality of the resulting clusters. We considered two categories of metrics: \textit{benchmark metrics}, where there is some ``ground truth'' labeling of latent topics or clusters against which to compare, and \textit{cluster metrics}, where we are comparing clusters (for instance, before and after a user action) against themselves.

When there are ground truth concepts available, the quality of clusters can be assessed via measures like the \textit{purity} and \textit{entropy} of a cluster~\cite{zhao2004empirical} (the degree to which a cluster contains only documents from a single ground truth topic, and the informational content of a topic with respect of ground truth topics, respectively). Chuang et al.~\cite{chuang2013topic} also measure cluster quality via a binary categorization procedure of mapping ground truth topics to clusters. We employ a similar mapping approach to generate standard measures like accuracy and precision.
%~\ac{We do something similar here for accuracy and precision}. 

Even when there are no ground truth labels available, it is still possible to measure the compactness, self-similarity, or distinctiveness of a particular cluster through distance metrics. Huang~\cite{huang2008similarity} explores a variety of different cluster distance metrics such as Jaccard Similarity, cosine similarity, and KL-Divergence as inputs to standard clustering systems, while Wang et al.~\cite{wang2017hierarchical} perform similar topic model-driven cluster analysis via the silhouette coefficient and the Cophenetic correlation coefficient.
%~\ac{Also silhouette score is useful here} 
A common pattern across these works is that differing distance metrics have differing patterns of performance across different text corpora, with no clear ``winner'' (a situation common across the problem of clustering in general~\cite{jain1999data}); we use these results to justify the collection of multiple related distance metrics where possible. We describe the subset of benchmark and cluster metrics we implemented in our simulation work in \S\ref{sec:pipeline-benchmark} and \S\ref{sec:pipeline-cluster}, respectively.

%This goes in methods I think:
%Ground truth topics may not be available. Or, if they are available, may have little to do with the latent space that might be constructed by algorithms like LDA. We wish to capture information about those cases as well. For unsupervised settings, we view the state of the model prior to a simulated action as the ground truth, and compute distances relative to that state.

\subsection{Sensitivity Analyses of Visual Analytics}

\begin{figure}
    \centering
    \includegraphics[width=0.5\textwidth]{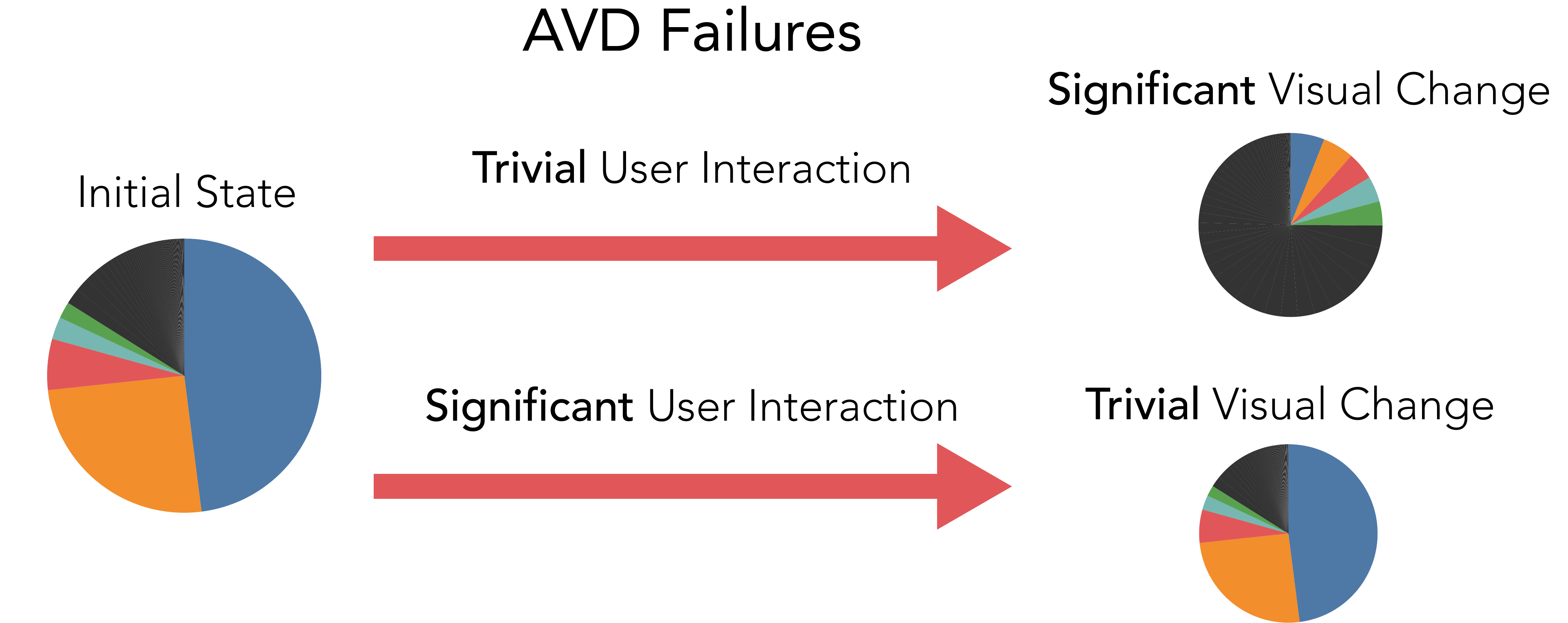}
    \caption{Examples of Algebraic Visualization Design ``failures,'' across a sample pie chart visualization of topic clusters in a corpus. If a user performs what they believe to be a trivial action (say, removing rare words) and it results in a large and uninterpretable shift in topic clusters, then the user may be unable or unwilling to trust or interpret the model. Likewise, if the user performs an action they intend to have large impact (such as dramatically changing the number of topics) and see little to no change in the model, they may feel a loss of trust or agency.}
    \label{fig:avd-failures}
    \Description{An initial pie chart of topic membership in a corpus contrasted with two resulting pie charts showing the corpus state after human actions that can indicate algebraic visualization failures. Either large visual change as a result of a trivial user action, or small visual change as a result of a significant user action.}
\end{figure}
There are many potential actions that an analyst can take over the course of analysis. For instance, an analyst looking at text data could plausibly make decisions about which texts to include or exclude from the corpus, whether or not to remove stop words or common words, and whether or not to stem or group words before the first chart is drawn or analysis is run. Each of these decisions could result in a dramatically different final analysis or visualization.  This ``garden of forking paths'' problem~\cite{pu2018garden} for visual analytics is often portrayed as an issue of reliability or replicability of findings. An emerging challenge in visual analytics is therefore how to capture and visualize the data flow that led to a particular experimental outcome~\cite{freire2012making}, or the visualize the robustness of a conclusion across a ``multiverse'' of different analytical paths~\cite{dragicevic2019increasing,kale2019decision,liu2020boba}. While the lack of exploration of data prep flows and hyperparameters (and its resulting impact on the model) has been identified as a ``troubling trend'' in machine learning generally~\cite{lipton2019troubling}, this issue of reliability across highly variable methods is particularly vital in text analytics, where existing statistical tools are often thought of as fragile or sensitive in the face of the semantic complexity of text corpora. For instance, Da~\cite{da2019computational} claims that for scholarship based on text analytics, ``what is robust is obvious... and what is not obvious is not robust'': that after one removes the scholastic conclusions from text analytics that are properties of idiosyncratic selections of datasets or choices of methods, the only remaining reliable conclusions are often so obvious or bland as to be uninteresting.

%\todo{AVD Figure? Or Does fig \ref{fig:exampleActions} cover these paths?}\ac{I think this is a result or interpretation of the result and it might be worth moving the AVD figure there.}

While our simulations are rooted in these sorts of sensitivity analyses, large changes in, say, accuracy or cluster coherence due to changes in the model are not intrinsically dangerous or user-unfriendly. Actions that result in large changes or sensitivities to the model might be perfectly acceptable if the user is aware that they are undertaking a disruptive action. We wanted to capture that sensitivity and reliability are contextual, and so we rely on the principles of Algebraic Visualization Design~\cite{kindlmann2014algebraic} (AVD). Following AVD, we hold that user expectations of the visualization of the model are met when small or large changes to the data result in commensurately small or large changes to the resulting output. In cases where there is an algebraic violation (say, an action that the user perceives as a minor adjustment to the text pipeline results in fundamentally different topics, or a user intends to induce a total reclassification of texts but merely shuffles around existing categories) as in \autoref{fig:avd-failures}, then the designers of topic modelling tools might wish to signal or otherwise alert the user to this mismatch.

Simulations (whether headless or based on particular interfaces) are especially useful in the AVD regime, as they allow the direct manipulation of inputs (data changes) and a direct measuring of outputs (visual or representational changes). For instance, Correll et al.~\cite{correll2018looks} use simulation of data quality issues to highlight visual designs that may not robustly or reliably surface important properties in distributions. More generally, McNutt et al.~\cite{mcnutt2020surfacing} propose the use of simulation results to automatically detect potentially misleading or unstable insights from visualization. An insight that is highly sensitive to particular conditions may not be reliable or generalizable. This work on using simulation to detect \textit{commensurate} changes in data and model output inform our methods.

%\mc{Would like to talk about headless testing here, but I'm not certain what the standard cites are in that space.}

%% file: text/3-methods.tex
\section{A Human-in-the-Loop Text Analytics Pipeline}\label{sec:methods}

%\begin{wrapfigure}{l}{0.5\linewidth}
%    \vspace{-4mm}
%    \centering
%    \includegraphics[width=\linewidth]{"./figs/LDA_pipeline"}
%    \caption{Overall Pipeline. Emoji with monocle icon are steps that can incorporate user input obtained via some visual interface. }
%    \label{fig:lda_pipeline}
%\end{wrapfigure}

In this section we describe the text analytics pipeline that we use to assign documents to topics and elicit user input (\autoref{fig:teaser}). First, we describe the overall text analytics pipelines, including the three types of metrics (benchmark, cluster, and topic) that we collect to assess the performance of the pipeline. Next, we describe how user actions are incorporated into the text analytics pipeline. We also describe the ways that we simulate these actions and compute their impact relative to an initial baseline run. 

\subsection{A Pipeline for Document Classification and Topic Elicitation}
\label{subsec:methods-pipeline}
We implemented a fairly standard pipeline that models documents as a bag of words and uses Latent Dirchlet Allocation (LDA) for topic elicitation and document classification. A single simulation `run' of the LDA pipeline represents some specific set of parameter configurations, for example, the number of topics provided to an LDA model or whether or not to stem tokens. \textbf{\textit{The baseline run} refers to the default pipeline and parameter configurations (tailored to a particular dataset, as described in our results), whereas all subsequent runs refer to some simulated user action.} We break down the steps of our pipeline into three stages: data preparation, model building, and performance assessment. %The classification of these stages is relevant as they dictate where and how possible user actions are executed within our pipeline.

\textbf{The data preparation stage} takes a corpus of text documents and creates of bag of words model for each document. A baseline run of the pipeline tokenizes and processes text into 1-grams, filtering out short or numeric tokens, removes stop terms, and finally stems tokens. We compute the term frequency (tf) and term frequency inverse document frequencies (tf-idf) statistics to create the document-term matrix for the model stage. In the baseline run we do not remove rare or ubiquitous terms by default.

\textbf{The model stage} trains a LDA Model with a set of default parameters. The default setting for the total number of topics is 10, but if the user has ground truth labels for their data then the total number of topics can be automatically derived from the label vector. Once the model is trained, we compute the topic assignment for documents. The LDA algorithm produces a posterior distribution of topic membership for each document and we use the \texttt{argmax} function to assign each document to one final topic. We can also extract the posterior distribution for the term-topic relationship.

Our last stage is \textbf{performance assessment}. We compute a set of \textit{benchmark}, \textit{cluster}, and \textit{topic} metrics. We provide more details on the calculation of these metrics in the subsequent subsections~\S\ref{sec:pipeline-benchmark} to~\S\ref{sec:pipeline-topic}, but provide a high-level overview here. Benchmark metrics compare the performance of the current run against some baseline. Benchmark metrics assess the accuracy and precision of the document classification. Cluster metrics how well documents are grouped together into clusters via assessments of cluster homogeneity, completeness, variance, and the document silhouette. Finally, topic metrics assess the distribution of terms across topics.

While topic quality measures are the norm in the visualization literature (see \S\ref{sec:related-topic-model-metrics}), studies have not previously examined the utility of benchmark and cluster metrics for examining or steering human-in-the-loop actions. These latter metrics depend on having some ground truth dataset to compare against and thus have been overlooked, however we see useful ways that these metrics can be used with or without an \textit{a priori} ground truth label.  When ground truth labels are known, we can compute both the magnitude and direction (i.e. improve model quality or not) of change introduced by a user action. When ground-truth labels are not available, it is possible to use the predicted labels from the  baseline run and to measure only the magnitude of change introduced by a user action. This approach is reasonable as prior research has shown that default parameters are highly influential in visualization design, often to the potential detriment of understanding~\cite{conti2005attacking,correll2018looks}. Users may not often change default model parameters and may use the initial classification results as a kind of default run.

This pipeline is implemented in Python using primarily the \texttt{scikit-learn}~\cite{scikit-learn} and \texttt{nltk}~\cite{ntlk} packages. The code is available online at \texttt{\url{https://osf.io/zgqaw}}. We have implemented the pipeline in a modular way that enables those that wish to expand on our approach to incorporate new user actions and even pipeline steps.

\subsubsection{Benchmark Metrics}\label{sec:pipeline-benchmark}
Given some document labels we compute benchmark metrics.
%With a supervised analytics approach, these initial labels would also be used to inform the model training procedures.
%and the predicted document classifications labels correspond directly to the initial set of labels. 
As LDA is an unsupervised method, and the relation between the topics generated by LDA and any a prior  document labels is not straightforward; it often up to the user to infer the semantic content or relation between topics.
%,  meaning that the predicted topic labels do not correspond to an document labels should they exist. As an example, an LDA model tuned to predict ten clusters will produce a set of topics with generic labels 0 to 9,
%When an initial set of labels exists, we attempt to discover the correspondence between predicted topic labels and the ground truth labels, which in turn allows us to compute accuracy, precision, and variance.  
We automatically derive this topic correspondence when computing accuracy and precision metrics. We define and calculate these metrics in a slightly different way here compared to supervised settings. Accuracy is a measure of how many documents with a common ground truth class are assigned to a common predicted topic. We provide details for this computation in Algorithm~\ref{alg:accuracy}. We compute the accuracy for each ground-truth class, as well as an average and weighted average for the entire run. Precision is a measure of how many documents of a common class are assigned to a common predicted topic and the purity of the predicted topic. We compute precision via the F-1 and Fowlkes-Mallows Index (FMI) metrics as shown in Algorithm~\ref{alg:precision}. 

\begin{algorithm}
\caption{Unsupervised Class Accuracy($D$,$T_g$,$T_p$)}
\label{alg:accuracy}
\begin{algorithmic}
\Require Documents $D$, Set of Ground Truth Labels $T_g$, Set of Predicted Topics $T_p$
\Ensure Class Accuracy
\State $A \leftarrow{} []$
\For {$T$ in $T_g$}
    \State Get subset of documents assigned to T
    \State \hspace{5mm}  $D_{t} = D \in T$
    \State Get predicted class with largest number of $D_t$:
    \State \hspace{5mm} $D_{t} \in T_p$
    \State Get class specific accuracy $A'$
    \State \hspace{5mm} $A' \leftarrow{} max(\frac{|D_{t} \in T_p|}{|D_{t}|})$
    \State $A.append(A')$
    %\comment{Number of documents belonging to ground truth label T}
    %\State $D_{T}^{'} \leftarrow []$
    %\State  $D_T  = |\{D_{T_g} \in T_g | D_{T_g} = T\}|$ 
    %\For{$P$ in $T_p$}%
    %\State $D_P  = |\{D_{T_p} \in T_p | D_{T_p} = P\}|$ %\Comment{Number of documents with label T classified to topic P}
    %\State Append $D_P$ to  $D_{T}^{'}$ 
    %\EndFor
    %\State $A' = max(D_{T}^{'})/D_T$%\Comment{Accuracy for a single ground truth label}
    %\State Append $A'$ to  $A$
\EndFor
\State \textbf{return}  $mean(A)$
\end{algorithmic}
\end{algorithm}

\begin{algorithm}
\caption{Unsupervised F1,FM1($D$,$T_g$,$T_p$)}\label{alg:precision}
\begin{algorithmic}[1]
\Require Documents $D$, Set of Ground Truth Labels $T_g$, Set of Predicted Topics $T_p$
\Ensure F1, FMI
\State $TP,TN,FP,FN =  0$
\While {$i \leq10000$}
\State Randomly select document pairs:
\State \hspace{5mm} $[D_1,D_2] =  sample(D,2)$
\State Get true and predicted topics for each document
\State \hspace{5mm} $t_{D_n}^g$,$t_{D_n}^p$ where $n \in (1,2)$
\State True Positive if pairs have the same $T_g$ and $T_p$
\State \hspace{5mm} $TP = TP +
    \begin{cases} 
     1,& \text{\textbf{if }} t_{D_1}^g = t_{D_2}^g  \text{ AND } t_{D_1}^p = t_{D_2}^p\\
     0
    \end{cases}$
\State False Negative if pairs have the same $T_g$ but different $T_p$
\State \hspace{5mm} $FN = FN +
    \begin{cases} 
     1,& \text{\textbf{if }} t_{D_1}^g = t_{D_2}^g  \text{ AND } t_{D_1}^p \neq t_{D_2}^p\\
     0
    \end{cases}$
\State False Positive if pairs have different $T_g$ but the same $T_p$
\State \hspace{5mm} $FP = FP +
    \begin{cases} 
     1,& \text{\textbf{if }} t_{D_1}^g \neq t_{D_2}^g  \text{ AND } t_{D_1}^p = t_{D_2}^p\\
     0
    \end{cases}$
\EndWhile
\State $F1 = TP / (TP + 0.5\times(FP+FN))$
\State $FMI = TP / \sqrt{(TP + FP) \times (TP + FN)}$
\State \textbf{return}  $F1, FMI$
%\State Randomly select two pairs of documents
%\State Compared ground truth (gt)  classes and predicted classes (pc)
%\State IF same gt for the pairs is the same AND pc for the pairs is %the same then TP +=1
%\State IF same gt for the pairs is the same AND pc for the pairs is %different then FN +=1
%\State Same kind for logic to compute FN and FP.
%\State Re-run 10,000 times (essentially get 10,000 random pairs)
%\State Compute and FM1 for the aggregated TP FP TN FN results
\end{algorithmic}
\end{algorithm}

\subsubsection{Cluster Metrics}\label{sec:pipeline-cluster}
%\ac{Benchmark and cluster metrics are pretty similar with respect that they assess cluster quality. Benchmark metrics assume a ground truth, cluster metrics do not. Need to make this clearer}\mc{Trying something here in related works to percolate down to here. Let me know if it works for you}
%\mc{Nope, this doesn't work for me. Reverting back.}
Cluster quality metrics evaluate how well documents are grouped together. We use the term \textit{cluster} to emphasize that the quality of the resulting topics, as expressed by the term distributions across topics, is not under consideration, only the grouping of documents. We compute overall quality of the classification via homogeneity, completeness, variance, and silhouette. The former three metrics require a document classification label, whereas the silhouette metric does not. Homogeneity is a score between 0 and 1 that measures how many predicted clusters contain data points of a single class; results closer to 1 indicate better homogeneity. Completeness is an overall assessment of whether all documents in some ground truth class belong to a single cluster: it also produces a score between 0 and 1. Variance is a harmonic mean between homogeneity and completeness. Homogeneity, completeness, and variance are computed at the level of each run only, rather than for each ground-truth class as can be done for accuracy and precision. The silhouette co-efficient is a measure of similarity of documents within a common cluster relative to documents in other clusters. The resulting silhouette co-efficient values ranges from -1 to 1, with a score of 1 meaning that a document is nearly identical to others within the same cluster. %We can also use the per document topic probability membership to assess the quality of cluster assignments. Across multiple runs, we can compare changes in silhouette scores, document membership probability, and the shift in the set of documents that are relevant across predicted topic clusters. 

\subsubsection{Topic Model Metrics \& Visual Analysis}\label{sec:pipeline-topic}

\begin{figure}
    \centering
    \includegraphics[width=0.45\textwidth]{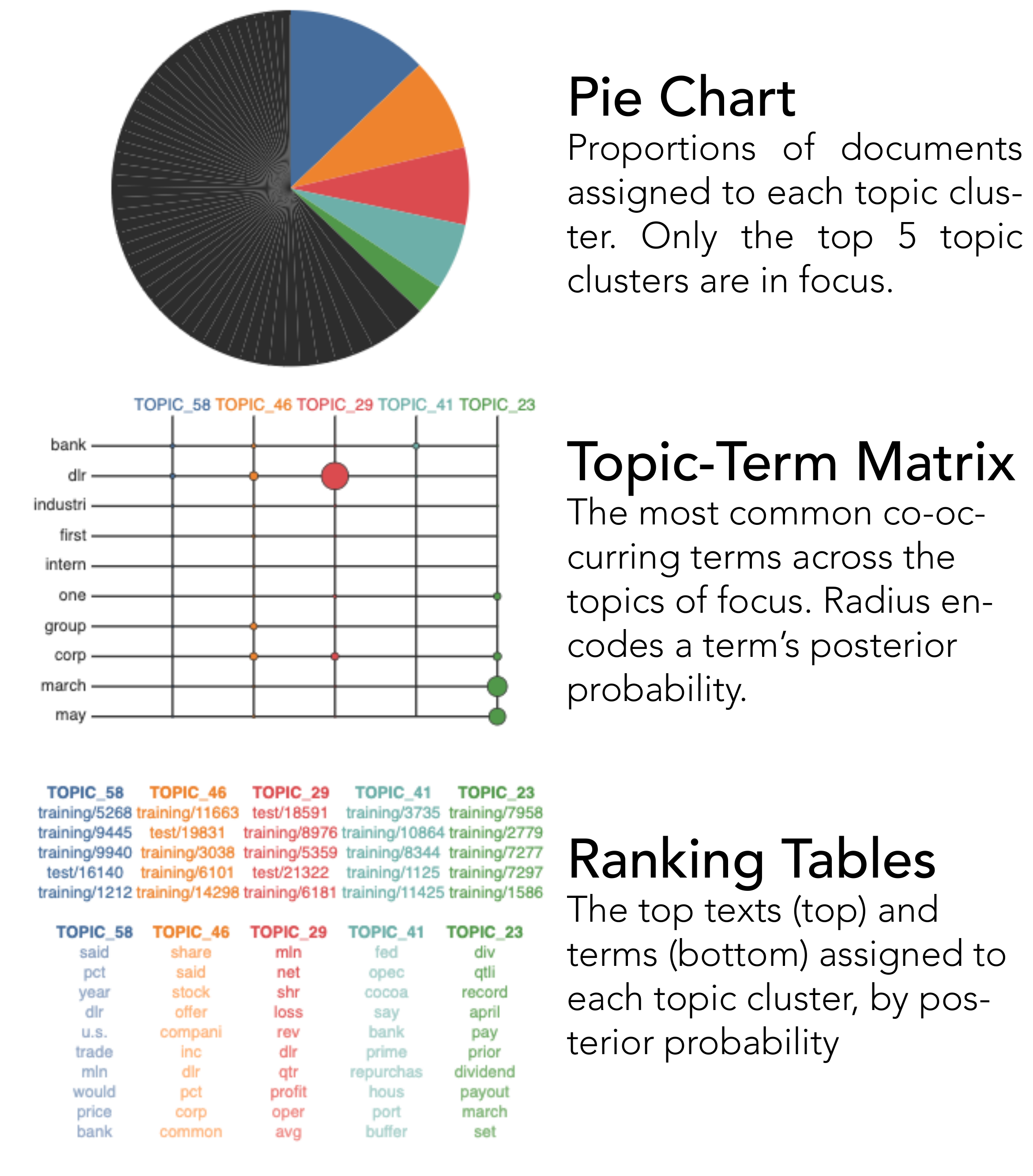}
    \caption{The views we include in our visual analyses of a topic model, representative of a hypothetical visual analytics system. Note that the viewer receives detailed information about only a few topics at a time.}
    \Description{The three views we created of the state of a corpus state with respect to the topic model: a pie chart of document membership in topics, a term-topic matrix of token loadings in topics, and ranking tables of the top N documents and tokens associated with a subset of topics.}
    \label{fig:vis-schematic}
\end{figure}

Topic quality metrics concern the term-topic relationship. A simple way to summarize this relationship is to observe the top terms across topics according to the posterior probability computed by LDA. We summarize the diversity of topics by computing the KL-divergence, Jensen-Shannon similarity, and cosine similarity. We excluded some of the more complex bespoke algorithms mentioned in \S\ref{sec:related-topic-model-metrics} for reasons of performance, complexity, or specificity to particular corpora or models (e.g. a particular word embedding model, or a dataset of human-generated responses). %We can compare topics \textit{within} runs (the homogeneity of the all the topics in a particular model), as well as \textit{across} runs (the formation across simulation runs as well across runs. The latter comparison allows use to surface whether similar topics emerge in response to user actions. 
%Due to limitations of space these similarity metrics are included in the online materials.

Under the assumption that only a subset of topic information will be visible in a particular view, we also perform a \textit{visual} comparison of topics through an exemplar topic dashboard, inspired by existing commonly used visualizations (see \S\ref{sec:related-topic-model-visualization}). \autoref{fig:vis-schematic} shows an example view constructed for this purpose. Rather than a fully-functional interactive system, these views are meant to act as ``fruit flies for visualization''~\cite{rensink2014prospects} and provide a useful proxy for illustrating the practical magnitude of changes in human terms. First, we use a pie chart to convey the number of documents contained in each topic cluster. Second, we show a term-topic matrix view (\autoref{fig:avd-failures}) of the distribution of the top ten terms across the five largest topic clusters. These top-ten terms are established by summing and ranking the occurrence of these terms across topics. We use a circle mark with a variable size to encode the posterior distribution for each term across topics. Finally, we show the list of top five documents and top ten terms per topic clusters. This visual analysis allows us to make claims related to the algebraic relationships between actions and visualizations (see \autoref{fig:avd-failures} for examples).

\subsection{User Actions}
\label{subsec:method-useraction}
At the heart of our simulation is a list of \textit{simulated user actions} that correspond to either analytical choices made prior to model generation, and/or attempts to ``steer'' a model towards a more useful state. Here we summarize the user actions that we anticipate would be commonly carried out through some kind of interactive user interface. We make no assumptions about the elements of the user interface or visual encodings that users would interact with to carry out an action, for example, adjusting a slider to increase or decrease the number of topics. Instead, we focus on the effect the user interface manipulations would have on an underlying text analysis pipeline relative to an initial set of results produced without that particular user input.  In \autoref{fig:teaser} we show where user actions are incorporated into our LDA pipeline, we elaborate on how these user actions are incorporated in the subsequent subsection. %A key component of understanding the impact of these user manipulations is to map a user's action to a specific stage of the analysis pipeline. %For example, adding or removing stop words is a user action that needs to be incorporated in data pre-processing stage of our pipeline, whereas updating the number of topics impacts the modelling phase of the pipeline. These considerations are important for informing when and how to update our underlying text analytics pipeline with new information from the user. We break down user actions according to effects on data processes, modelling, or performance assessment.

\subsubsection{Pipeline Updating and Retraining Assumptions}
We make the assumption that each individual user action, as opposed to a set of user action taken in sequence, triggers the rerunning of the corresponding pipeline step. For example, adding new stop terms requires running the text analysis pipeline from data pre-processing forward, whereas joining clusters together only re-computes the metrics but does not re-run the entire pipeline.
%We make this assumption because we acknowledge the different systems will choose to incorporate user actions in different ways, for example the user might need to click a ``re-analysis'' button to update their initial results after perform one or more actions on the user interface to adjust pipeline and model parameters. 
We consider the choice of when and how to update a machine learning pipeline to be a design decision that is complimentary but broader in scope than our current research that seeks to explore nuanced user actions in detail. Moreover, this design choice can serve as a confound in our analysis because it obscures \textit{which} action, or type of user action, had the largest impact on changing the result. That being said, our present simulation approach can be expanded to support inquiries about updating or retraining procedures in the future. 

\subsubsection{Refinement Implementation Choices}
\label{sec:action_choices}
%\ac{Thinking extension and implementations are part of the user refinement.} 
%The user refinements that we have implemented reflect a sample of only possible topic model refinements that a user could or would want to apply. 

There are many potential user actions one could envision impacting a topic model. Lee et al.~\cite{lee2017human}, in an interview with topic model users, identify a list commonly requested actions such as adding or removing words, removing documents, and splitting topics. Building on this work, Smith et al.~\cite{smith2018} propose a list of actions for human in the loop topic models. While we use this list to motivate our implemented refinements, and attempt to maintain at least the spirit of these requested actions, a limitation of this prior work was that it only considered the types of refinements that users wanted to engage with, and did not go further to determine how those actions should be instantiated in a machine learning process. For example, when a user wishes to remove a word from a topic (the modal requested topic model refinement), how should LDA respond? There are several ways. E.g., we could add the word to our list of stop terms or we could  modify the probabilities for the topic-term distribution; in both circumstances we would have to retrain the model from scratch. Another alternative is to avoid retraining the model and instead provide only superficial changes so that the user has a sense of control but the topic model does not meaningfully incorporate the modifications-- many topic model systems do take this approach. Similarly, there are many actions that could have aesthetic impact (such as reordering or relabeling topics, or hiding irrelevant topics), and ought to be considered from the standpoint of the UX of a human-in-the-loop topic modeller, and could potentially impact the perceived trustworthiness or utility of a particular topic model, but that would not impact the underlying topic model structure in any concrete way. 

Our work attempts to bridge the gap between users' desired refinements and the ways that they could be practically incorporated into a machine learning regimen. More recent work by Kumar et al.~\cite{kumar2019didn} does begin to explore how user actions could be incorporated into topic model priors, but we believe that priors can have only so much influence on a model outcome, and that there are potential areas of impact across the entire LDA pipeline, rather than just manipulation of priors. We focus on different places in the LDA pipeline where we thought incorporation of user actions made sense, from preparation and modelling to performance assessment, and we similarly made decisions about when the LDA \textit{pipeline} (not just the model) should be rerun. Through this process we found that, in many ways, machine learning models are not well equipped to incorporate user input, as the language of hyper parameters does not necessarily map directly to user intent: different kinds of topic modelling algorithms can result in different affordances for user interaction and modification at different stages. Our decision to implement a standard LDA pipeline closed of some of these avenues (for instance, a hierarchical topic model would present a different notion of merging or splitting of topics than our pipeline). While we acknowledge the limitations of our particular pipeline as instantiated, we have attempted to craft our pipeline to be as modular and extensible as possible to afford experimentation with other kinds of actions and algorithms. 

%We acknowledge that the decisions we made about when and where to incorporate user refinements representations both an advancement over prior work but also a limitation. Being aware of this limitation fairly early in our research, we also tried to craft our pipeline in such a way that it could be possible to add refinements and extensions at a later date.

%We used these lists as guides, but our choice of a more traditional algorithm like LDA as opposed to more flexible methods constrains the actions we could implement. For instance, while Lee et al. report that adding and removing individual words was the modal requested topic model refinement, the standard implementation of LDA does not allow for the post-hoc removal of individual tokens. We have attempted, where possible, to preserve the spirit of these actions (for instance, the exclusion of rare or ubiquitous) while keeping to our LDA implementation.

%Similarly, there are many actions which could have aesthetic impact (such as reordering or relabeling topics, or hiding irrelevent topics), and ought to be considered from the standpoint of the UX of a human-in-the-loop topic modeller, and could potentially impact the perceived trustworthiness or utility of a particular topic model but that would not impact the underlying topic model structure in any concrete way. These aesthetic actions were also not considered by our simulation.

\subsection{Simulating User Actions}
\label{subsec:methods_sim}
As per our rationale in \autoref{sec:action_choices}, we build upon a set of interactions for topic modelling proposed by Smith et. al.~\cite{smith2018} and categorize actions according to their impact our text analytics pipelines. In~\autoref{fig:teaser} we summarize the user actions that we investigate.

\textbf{Preparation-related user actions} are those that trigger a re-start of the entire pipeline because they fundamentally change the distribution of terms or texts. These actions include:
\begin{itemize}
    \item The choice to remove stop terms
    \item Adding or removing stop terms to an existing list
    \item The choice to stem terms or not
    \item Removing rare words or ubiquitous words via an occurrence threshold
    \item Removing texts from the document corpus
\end{itemize}

These actions are interpreted in different ways by our LDA pipeline. The decision to remove stop terms or to stem terms are binary yes/no decisions that primarily impact which terms are used by the model, as well as the distribution of these terms across documents. When choosing to remove stop terms, a user can also exclude terms from a default list of English stop terms or they can add terms to an existing list. To simulate these actions, we carry out approximately 30 iterations where  a random number of stop terms are either excluded or included relative to the default list. Removing rare or ubiquitous terms requires the user to supply a numeric threshold value ranging between 0 to 100\%. For removing rare terms, we defined a set of thresholds (0.01\%, 1\%, 2.5\%, 5\%, and 10\%) where any term with a term frequency \textit{less than} the threshold value is removed. For removing rare terms, we defined set of threshold values (99\%, 95\%, 90\%, 75\%, 60\%, 50\%) where any term with a term frequency \textit{greater than} the threshold is removed. Finally, a user can choose to remove documents from a corpus, for example if they find some texts irrelevant and don't wish them to be considered when constructing the LDA model. We simulate this action by specifying a percentage of documents(5\%, 20\%, 25\%, 40\%,50\%) to remove from the corpus.  For these last three types of user actions we selected a fixed set of thresholds in lieu of sampling because it allows us to more efficiently explore the space of possible and reasonable user choices.  

\textbf{Model-related user actions} require a retraining of the LDA model; this impacts the final document-topic distributions as well as term-topic distribution. The most salient parameter to LDA is the number of final topics to generate. We generate a distribution of potential topic numbers that ranges from 2 to at most 25\% of the total number of documents in a corpus (or a maximum of 100 clusters, whichever is smaller). We sample uniformly from this distribution 30 times to simulate a user action of modifying the LDA parameters. 

\textbf{Assessment-related user actions} are those that modify the document-topic and term-topic distribution but do not require a retraining of the LDA model. One such action we simulate is splitting a single topic into two sub-clusters. We randomly select at most 30 cluster to split in two. A user may also wish to merge one or more cluster together. To simulation this action, we randomly select a total of N topic clusters (where $N\in(2,10)$) to merge together. After clusters are split or merged, we not only modify the predicted labels for documents but the topic membership probabilities as well. These modifications are used to recompute the benchmark, topic, and cluster metrics. 

\subsubsection{Data Collection and Analysis}\label{sec:methods-analysis}
Each simulation run produces data pertaining the run, documents, predicted topics, and ground truth labels (where present). For runs, we capture the specific user action, its impact, and the resulting overall benchmark, cluster, and topic metrics. For each document we capture its probability of assignment to a topic in addition to its final topic assignment; for simplicity we only output document-topic membership probabilities that are greater than 0.001. For each topic we output the top 100 documents and terms and their probability of assignment to each topic. We use this data to compare performance across runs, but also to compare topics within a simulation run.

We use descriptive statistics to summarize the changes in benchmark, cluster, and topic metrics over time. Finally, we compute a summary of a run $r$'s impact compared to a baseline run $b$, $S_{r}$, as the normalized $\ell_1$ distance between the two runs across our $M=8$ metrics:

\begin{equation}
\label{eqn:avg_dev}
   S_{r} = \frac{\sum_{i=1}^{M}|b_{i} - r_{i}|}{M}
\end{equation}

%where $i \in (0,M)$ and $M = 8$, the total number of benchmark and cluster quality metrics we evaluate. We compute the difference between baseline, $b_i$, and simulation, $r_i$ runs for each metric. 
We use the final value of $S_{r}$ to rank the overall impact of user actions across our simulation runs. $S_{r} \in (0,1)$, where $S_{r} = 0$ indicates that the result is identical to the baseline run across all of our metrics.\textbf{ The data and analysis results are available in our online repository: \texttt{\url{https://osf.io/zgqaw/}}}. 

\subsubsection{Further Extensions \& Improvements}
We report on a very simple model of user action in this paper, both to avoid a combinatorial explosion of data but also to allow a fully automated simulation pipeline. For instance, it is likely that users will engage in a sequence of multiple actions upon being given a topic model in an initial state rather than just a single action as in our current reported data. While our pipeline does support simulation of concatenated actions, building up a coherent and computationally tractable way of modelling and reporting on an entire multiverse of different pipelines of concatenated user actions remains future work. Similarly, many of our simulated actions are applied to random topics or texts. User actions are likely based on both model properties (for instance, being more likely to split a larger topic cluster than a smaller one) and domain knowledge (for instance, having an ontology in mind and altering the topic clusters to fit this ontology). modelling the \textit{likeliest} user actions is likewise an area out of the scope of this paper (in Kumar et al.~\cite{kucher2015text} it requires an explicit modelling of user priors, as opposed to the random preferences of ``bad users''), and would require followup analyses of log data and specific user goals that are likely high dependent on context.

%% file: text/4-results.tex
\section{Results}\label{sec:results}
In this section we describe the application of our text analytics pipeline with the Reuters-21578 dataset that is widely used in the machine learning literature. Due to limitations of space, we relegate the analysis of additional datasets to the online materials; we briefly summarize key findings from these data in~\S\ref{sec:result-added}. %The second is a dataset of COVID-19 research articles released by the Allen Institute, that have no \textit{a priori} labels.  We describe the characteristics of these two datasets and summarize the effect of different types of user action as they are measures through benchmark, cluster, and topic metrics.

\subsection{Reuters-21578}\label{subsec:results-reuters}
\subsubsection{Dataset Description}
\begin{figure}[t]
    \centering
    \includegraphics[width=0.5\textwidth]{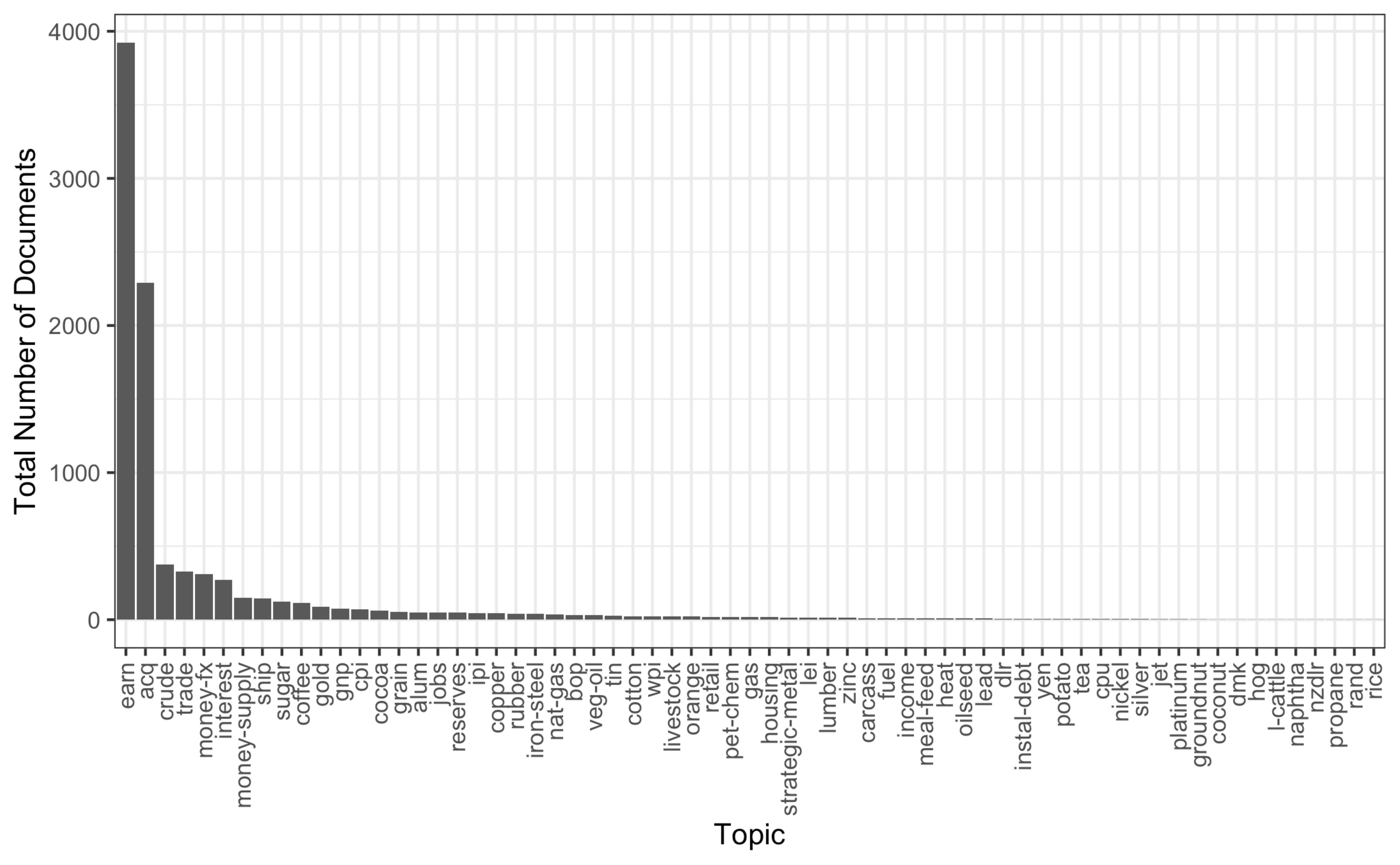}
    \caption{Ground truth topic distribution across the Reuters-21578 dataset.}
    \Description{A histogram of which documents are associated with which ground truth topics in the Reuters-21578 dataset. The distribution is extremely skewed and long-tailed: nearly 4000 documents are associated with the most popular topic, but most of the other topics have far fewer associated documents.}
    \label{fig:gt_dist}
\end{figure}
The Reuters-21578 dataset is routinely used as a benchmark for text categorization algorithms~\cite{lewis1997reuters}.  The dataset comprises 10,788 documents that have been manually assigned to one or more of a possible set of 90 topics. We limit our analysis to documents that have only one topic assigned to them, which is 9,160 (84\%) of all documents.
%that retain the 65 of the original 90 topics. 
We use this set of labels as our ground truth to assess the performance of our unsupervised text analytics pipeline. The distribution of documents (\autoref{fig:gt_dist}) across the ground truth topics varies from as few as a single document per topic to a maximum of 3,923 documents per topic.

\subsubsection{Dataset Specific Pipeline Optimization}
In this study we have attempted to calibrate the initial set of pipeline parameters to the dataset rather than rely on na\"ive defaults. For example, the \texttt{scikit-learn} default for the total number of topics to produce is 10 and the token vectorizer similarly has a set of defaults. We modify these default parameters through a combination of a prior knowledge and experimentation in order to generate our baseline results. We use the Reuters-21578 dataset to primarily showcase our results: in this dataset we know the actual number of ground truth topics and so set the LDA parameters accordingly (although we experimented with other possible parameters at other stages of our pipeline). 

We used the benchmark metrics, described in~\autoref{sec:pipeline-benchmark} to guide our calibration process. The full compliment of parameter settings and other considerations are available in our online materials.  We refer to this dataset-calibrated pipeline as the baseline run. Importantly, our objective in calibrating the LDA pipeline was not to create a perfectly tuned algorithm, but instead a reasonable baseline from which we measured the impact of user actions; we argue that it was useful to leave room for a user actions to potentially further optimize this baseline.

\subsubsection{Benchmark and Cluster Metrics Capture Different Effects of User Actions}

\begin{figure*}[t]
    \centering
    \includegraphics[width=\textwidth]{"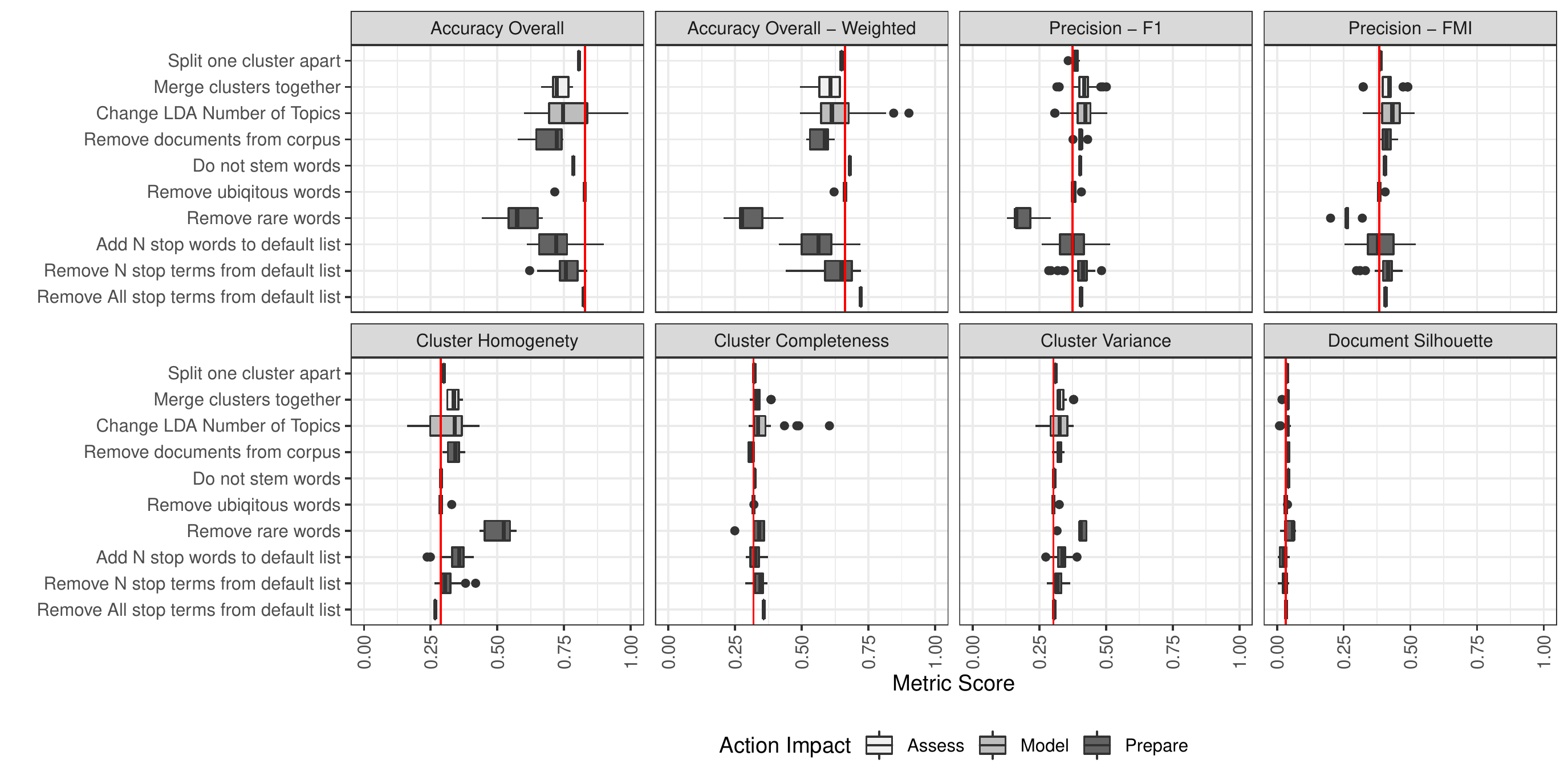"}
    \caption{Variations in benchmark metrics over different runs. We show the distribution of results across multiple simulation runs with different parameter configurations that a user could set. All metrics produce a value between 0 and 1, with the exception of silhouette, whose theoretical output range is -1 to 1, although only of the runs produced a value slightly below 0. The red line indicates the performance of the baseline run.}
    \Description{Box plots showing the impact of our simulated actions on our quantitative quality metrics. For most metrics the impact is small, but some actions like removing rare words or documents appear to result in consistent decreases in accuracy.}
    \label{fig:benchmark}
\end{figure*}

In~\autoref{fig:benchmark} we show the distribution of results from 167 simulation runs, each with a different potential user action. We show the performance of the default run as a red line; the extent of deviation from this red line shows how much of an impact a potential user's change has.

\begin{figure}
    \centering
    \includegraphics[width=\columnwidth]{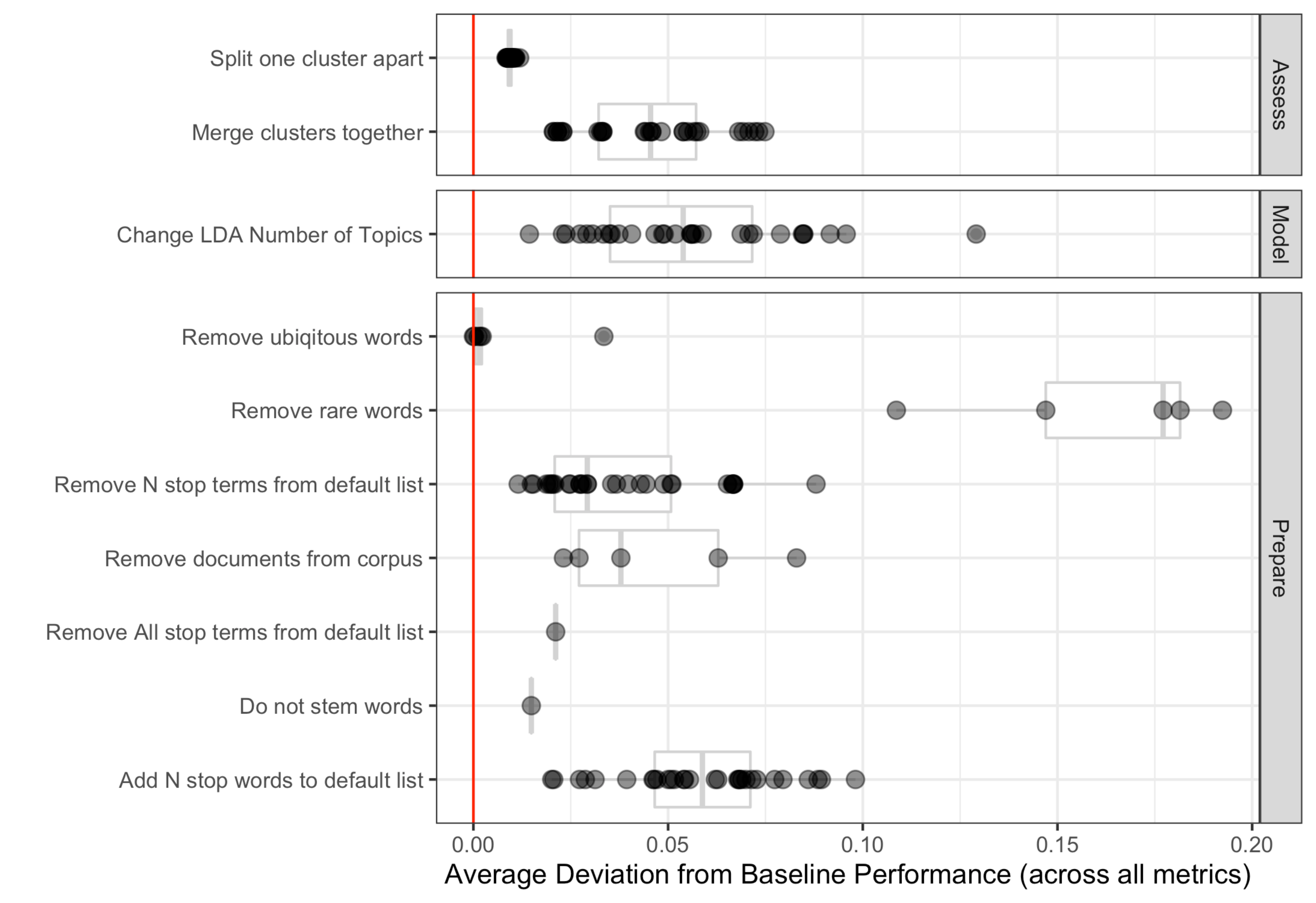}
    \caption{The impact of a simulated user action $S_{r}$ relative to a baseline run. Each point represents a single simulation run. Some types of actions have more runs associated with them because the possible space of parameters to sample from is larger compared to others. The red line is at zero and indicates near identical results compared to the baseline run.}
    \Description{Dot plots with boxplot overlay of the impact of user actions as defined by a weighted sum of all of our existing quantitative metrics. Removing rare words had the most consistent holistic impact, while others like splitting clusters and removing ubiquitous words had very little impact.}
    \label{fig:action_impact}
\end{figure}

First, we observe that different individual metrics measure the degree of change differently. Accuracy appears to be the most sensitive compared to other metrics that vary less. A reminder that the calculation of accuracy here is not identical to the calculation of accuracy in supervised settings, but instead a measure of how many documents with a ground truth class appear in a common cluster (see~\S\ref{sec:pipeline-benchmark}). The calculation of accuracy is thus tightly coupled to the size and composition of different cluster, whereas other metrics are more robust across different cluster distributions and sizes. A large deviation in accuracy is indicative that there are likely large changes in cluster membership, which may or may not be reflected by the more global qualities of clusters measured by the other metrics. We can summarize these changes holistically according to an action impact score, $S_{r}$ (see~\S\ref{sec:methods-analysis}) and by directly visualizing how different models classify documents (\autoref{fig:action_impact}). By ranking the resulting $S_{r}$ values, we establish that removing rare terms is one of the most impactful (or disruptive) actions a user can take, whereas removing ubiquitous terms is one of the least impactful. This finding was not altogether surprising: the distribution of terms across documents in generally sparse and so long as the threshold for ubiquitous terms remains above some reasonable level, this action is unlikely to result in the removal of all that many terms by count. Removing rare terms is more impactful not only because the matrix is already sparse but because it will substantially change the feature space of the LDA model. Surprisingly, changing the number of topics did not result in as large of change as anticipated. In order to see a large impact, the user would need to configure LDA to run with substantially different parameters, for example as few as 7 when the we know there are roughly 65 topics. We suspect that this relative insensitivity is because changing the number of topics does not alter the feature space, only the distribution of those features in topics. 

\begin{figure*}
    \centering
    \includegraphics[width=0.95\textwidth]{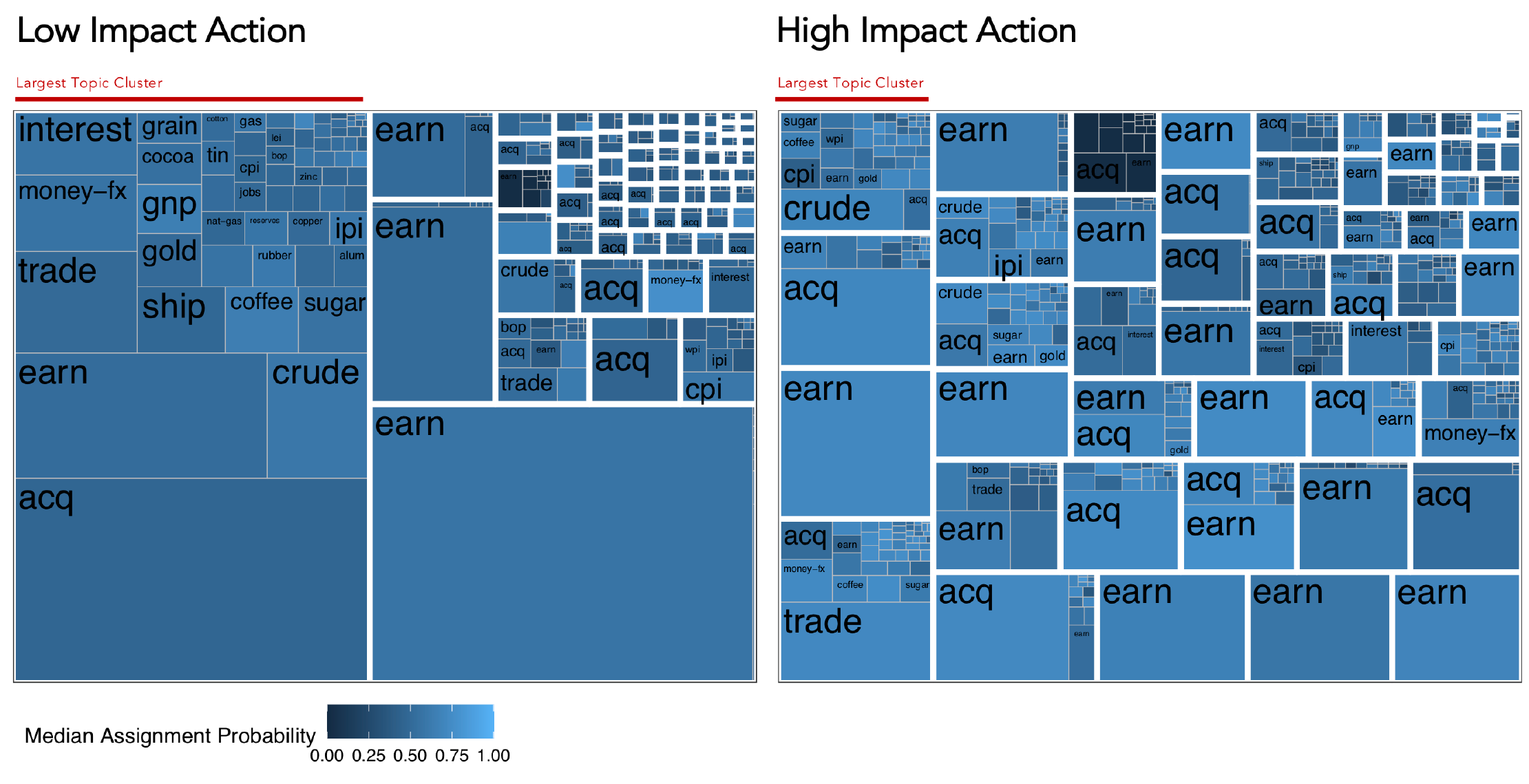}
    \caption{A nested tree map show the composition of ground-truth classes with predicted topics. We compare between the simulations runs with the lowest (A) and highest (B) impacts across all of our performance and quality metrics. Predicted topics are separated by thick white lines and within each predicted topic we nest a treemap that shows the breakdown ground-truth labels assigned to the predicted topics. The color of the blocks within this nested treemap shows the average probability of document membership.}
    \Description{Two nested treemaps of high impact and low impact actions. The low impact action maintains a structure of one dominant topic containing a long-tailed distribution of ground truth topics. The high impact action ha a more uniform distribution of documents across topics, with no one topic clearly dominating.}
    \label{fig:gt_compared}
\end{figure*}

In~\autoref{fig:gt_compared} we examine how documents in different ground-truth topics are assigned to predicted topics within the highest and lowest impact ($S_{r}$) simulation run.  We use a nested tree map which shows the predicted topics (separated by thick white lines) and with those the ground truth composition. First, it is noticeable that the low impact user action(~\autoref{fig:gt_compared}A) allocates many documents into a single predicted topic whereas the high impact user action of removing rare terms results in many smaller topic clusters. The second difference is that average cluster membership probability, indicated by color, is generally higher amongst the high impact run compared to the low impact results.

While our tree map visualizations are very different, they do not necessarily indicate an improvement in the overall cluster quality; they're both potentially sub-optimal results but for different reasons. The properties of the dataset are a strong indicator for why this is. The Reuters dataset is imbalanced, with two comparatively large topic clusters. The influence of these two dominant clusters is difficult to impact using only a limited set of parameters afforded to them. However, it may not be immediately obvious to the user that this is a desirable thing to do.  Moreover, these findings strongly suggest that the impact of user actions is also dependent on the characteristics of the dataset itself. This observation provokes us to reflect on our own pipeline and assess whether we have given users the necessary levers and waypoints to make meaningful and impactful through their human-in-the-loop interactions. 

We include additional analyses of our metrics, including topic-based metrics such as KL-divergence, Jensen-Shannon similarity, and cosine similarity, in our online materials.

\subsection{Visual Analysis}
%\ac{I collapsed topic model metrics and visual analysis into one section. There was just so much overlapping content between the two that it made sense to make them one thing}.
While our simulation focused on how user actions impact holistic measures of topic model utility and interpretability, in many practical systems the analyst may only have access to a ``snapshot'' of the topic model at a particular instance. These visualizations often either present an overview of the entire corpus in terms of topic membership or detailed per-topic information. The former is limited in the amount of detailed changes that can be noticed (an analyst may not notice small changes in proportions of documents belonging to particular topic clusters), and the latter is limited by the amount of the complexity that can be shown at once: term-topic matrices, for instance, often only show a small number of ``top'' tokens or topics, as it is infeasible to provide information about tens of thousands of tokens in detail in one view, and rely on interactivity or different ordering metrics to opportunistically surface different parts of the dataset~\cite{alexander_serendip_2014}. These abstractions and summarizations can result in the potential for AVD ``failures''~\cite{kindlmann2014algebraic} (see \autoref{fig:avd-failures}) or visualization ``mirages''~\cite{mcnutt2020surfacing}, where either important updates to the model fail to be represented in a salient way in the resulting visualization, or the visualization of a model may be highly altered visually without much change to the underlying topics or classification accuracy.

\begin{figure*}
    \centering
    \includegraphics[width=\textwidth]{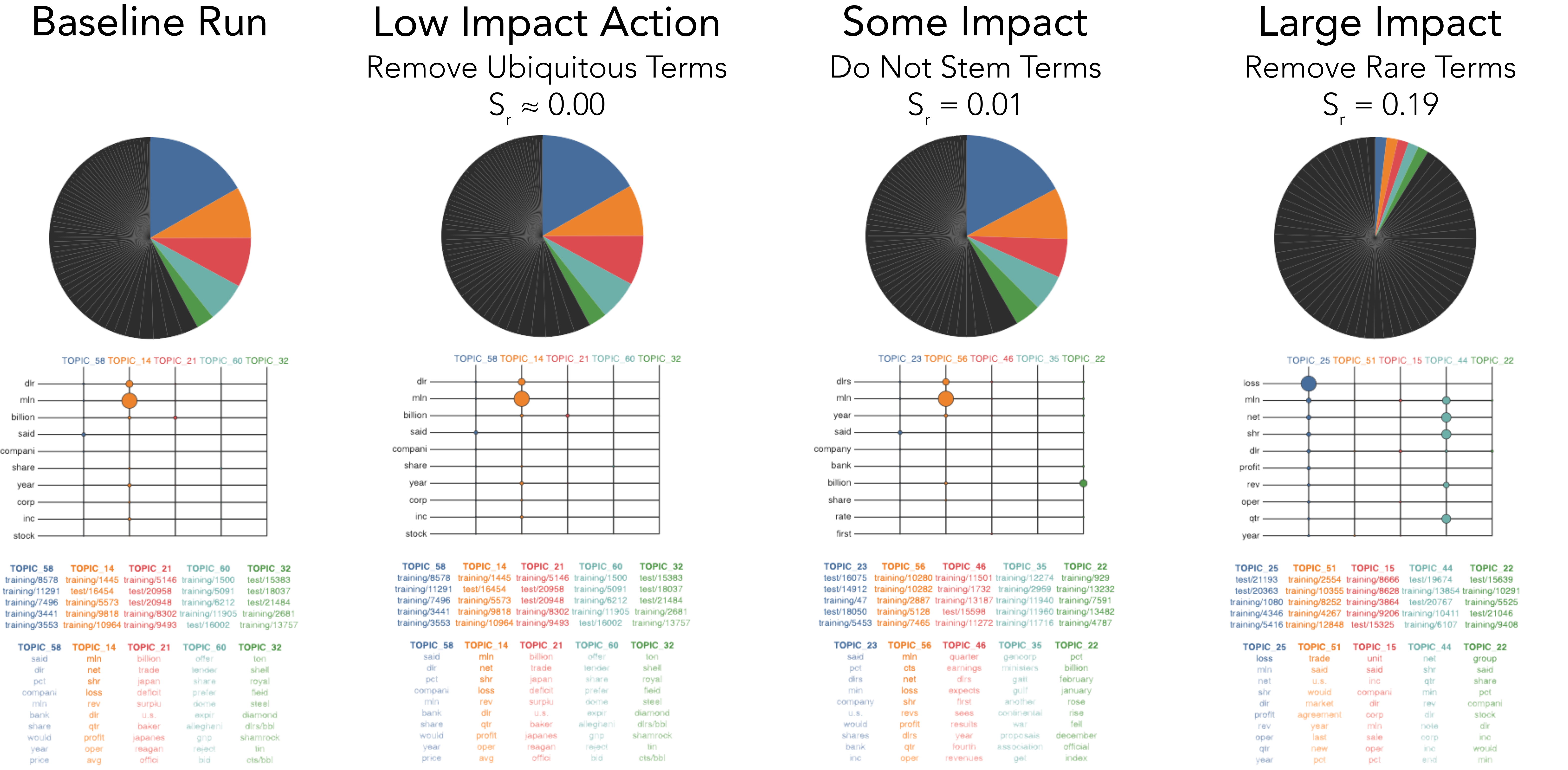}
    \caption{Under the Algebraic Visualization Design framework~\protect\cite{kindlmann2014algebraic}, existing visualizations of topic models may not adequately capture impactful changes, as shown in these examples of sample visual outputs of topic models before and after a simulated user action. As visualizations of topic models may only provide information about a few top topics or tokens, some actions may have almost no visible impact on the model. Other actions may have only minor impact (for instance, a reordering of top tokens or relabeling of topics). Only a select few of the actions we considered, such as removing rare words, resulted in large visual changes, although the specific actions that were most ``disruptive'' appears to be corpus dependent.}
    \Description{Four examples of visualizations of topic models: a baseline run before any actions are taken, and low impact, some impact, and large impact actions. The low impact action looks nearly visually identical to the baseline, whereas the large impact action has many differences, such as more uniform topic allocations and very different top terms and topics.}
    \label{fig:topic_qual}
\end{figure*}

In Figure \ref{fig:topic_qual} we show how high- and low-impact actions might be represented by a simple visualization system with a radial view of topic membership and a term-by-topic matrix view, both common summaries for individual topic models (\S\ref{sec:related-topic-model-visualization}). We use the $S_{r}$ score to select user actions that have low, middling, and high impact on the results of the topic modelling and document classification.

We visualize this impact through a sample design meant to emulate views that are common in standard topic model visualization systems (see \S\ref{sec:pipeline-topic} for more details). The lowest impact simulation is a user action that removes ubiquitous terms that occur in more that 75\% of documents in a corpus, with an $S_{r}$ value very close to $0$. Forgoing token stemming produces some impact, with $S_{r}=0.01$. The user action with the largest impact is increasing the threshold at which rare terms are removed, with $S_{r} = 0.19$; in this simulated action, the user opts to remove terms that occur in fewer than 10\% of documents.

Comparing these actions to a baseline run we see that many visual and textual components of the summaries are highly variable across runs: for instance, the actual numeric label of a topic, e.g. ``\texttt{TOPIC\_$n$}'' is likely to change even if the identity of the topic in terms of posterior probabilities across documents or terms is similar. Likewise, since the posterior probabilities in the (sparse) term-topic matrix are quite small and variable, the actual subset of the matrix that is visualized can vary wildly. A selection criteria that takes into account ranking or joint probabilities (as in our case, where terms are selected for inclusion in the matrix based on the extent to which they appear in the top 100 most probable words across all of the top $n$ topics of interest) can result in reordered, mismatched, or even entirely disjoint sets of corpus-wide ``top'' words across runs, even between actions that do not otherwise create large difference in performance or text classifications. 

However, other parts of the visualization remain unchanged unless dramatic changes to the model occur. For instance, the top words of individual top topics often were very similar across runs, perhaps changing order but not identity. Only extreme actions unreflective of common practice (such as deleting large percentages of rare words) were sufficient to produce large changes in top tokens for our large topics. Similarly, the overall distribution of documents is often qualitatively similar (a dominant topic and then a long tail of rarer topics) no matter the action simulated. Many of our actions (such as reducing the number of topics, or removing ubiquitous words, or merging topics together) were more likely to impact the tails of this assignment distribution, and so might be subtle or invisible in a visual, corpus-level overview.

%From~\autoref{fig:topic_qual} we see that this minor changes results in no visual differences compared to the default run. ; this difference is reflected by minor changes the matrix view and pie chart view, but with larger changes in the document and term lists. 
%The pie chart shows that the number of documents classified to the topic clusters is roughly the same as the default run. 
%The matrix view shows that relative to the baseline run the top ten terms across the topics of interest are very nearly the same and the probability distributions of those terms across topics is also very similar. The document list appears to show a larger change, with the top five documents being very different compared to the default run. However, the topic term lists indicate that the two largest clusters are roughly the same as the default run, but that forgoing token stemming appears to change the composition of the subsequent smaller clusters. Finally,  Commensurate with~\autoref{fig:gt_compared}, the pie chart in~\autoref{fig:topic_qual} reflects that the topic model outputs classified documents into many smaller clusters compared to the default run that is dominated by few large clusters. The matrix views as well as the document and topic lists are also very different compared to the default run.

Overall, we see that even a rudimentary summary of user impact, such as our $S_{r}$ score, surfaces a range of possible states in a human-in-the-loop process. Knowledge of these states allows both users and designers to consider resulting visual changes for cluster (\autoref{fig:gt_compared}) and topic (\autoref{fig:topic_qual}) quality. %, barring nearly ubiquitous visually trivial changes such as the textual labels of particular topics, or the ordering of top terms within a list. 
It would be fruitful future work to further evaluate how users interpret the impact of their own actions, and to what extent their perceptions aligns with existing performance and quality metrics. Our study here sets the foundation for such research by constructing the means to surface to users the impact of their actions.

\subsection{Additional Datasets}\label{sec:result-added}
In the online materials we produce a similar set of results using the the COVID-19 Open Research Dataset Challenge (CORD-19). CORD-16 is a set of research publications made available by the Allen Institute for AI in partnership with other non-profit, academic, and governmental organizations. Initially released in March of 2020 it has been updated multiple times and has grown to encompass more than 200,000 documents with over 100,000 full texts. Here we examine a subset of 10,000 documents considering only titles and abstracts to make the analysis comparable to the Reuters-21578 dataset. Unlike the Reuters-21578 dataset, the CORD-19 dataset does not have any ground truth labels. Instead, we use these data to demonstrate how it is possible to use the predicted topic results from the baseline run as a ground truth to evaluate subsequent user actions. We show that the $S_{r}$, which measures the magnitude of deviations from some baseline default run, is informative irrespective of whether an \textit{a priori} ground truth labels exist. Indeed, the primary advantage of having a ground truth label is that is allows us to make judgements about the direction of change as well, specifically whether the overall model was improved or not. We did see in the Reuters-21578 that many of the user actions we simulated had a negative impact of model performance, but we could not make such an assessment for the CORD-19 data. When we consider individual user actions, we found that in the CORD-19 dataset removing ubiquitous terms had the largest impact with respect to the baseline run performance. We suspect this is because the CORD-19 dataset has much more related content compared to the documents in the Reuters-21578 dataset. However, the results from the CORD-19 confirm that user actions directly impact the feature space, which is part of data preparation, had large impacts.  We also found that user actions appear to provoke larger changes in the CORD-19 dataset compared to the Reuters data. These findings underscore how important the characteristics of the initial dataset are not only on model performance but on the impact of user actions. Moreover, they demonstrate the utility of even a rudimentary metric like our $S_{r}$ score to surface these differences. 
%\mc{I think this section could be either lengthened considerably (or at least mention that the actions that had the biggest impact were consistent/inconsistent with Reuters) or dramatically shortened (just mention that we can extend the methodology to other data sets and point to the repo, perhaps only in the limitations/conclusions}\ac{ended up lengthen it a bit more}.

%% file: text/5-discussion.tex
\section{Discussion}

We condense our findings down to a set of three important results:

The \textbf{outsized influence of pre-processing steps} in the resulting topic model. Many existing human-in-the-loop systems focus on manipulating a model once it has been generated (by e.g. adjusting clusters or providing classification feedback). However, the changes we observed from these sorts of manipulations were often times limited, whereas actions impacting the data model and the textual inputs into the topic model were much more influential.

The often \textbf{subtle or poorly predicted impact of actions} on the resulting topics, especially as they are commonly visualized with only a handful of topics or tokens ``in focus'' in a given time. Often actions would have to be extreme in degree (beyond what we would expect from ``reasonable'' tweaking of parameters) to reliably produce impacts that were visually apparent, while others would reliably produce large changes even at the lower levels of the parameters we tested.

The \textbf{damaging impact of unprincipled actions} on our various metrics related to accuracy and coherence. While many of our actions had \textit{impacts} on our various quality and performance metrics, in nearly all cases this impact was \textit{negative}. This points to either the reasonableness of our ``default'' parameters, or, perhaps more likely given the differing ``defaults'' we have observed in other topic modelling work and the unconstrained nature of our simulated actions, that random actions not motivated by a specific observed deficiency in the model are unlikely to have positive outcomes. 

\subsection{Implications for Design}

\begin{figure*}[htb]
    \centering
    \includegraphics[width=0.65\textwidth]{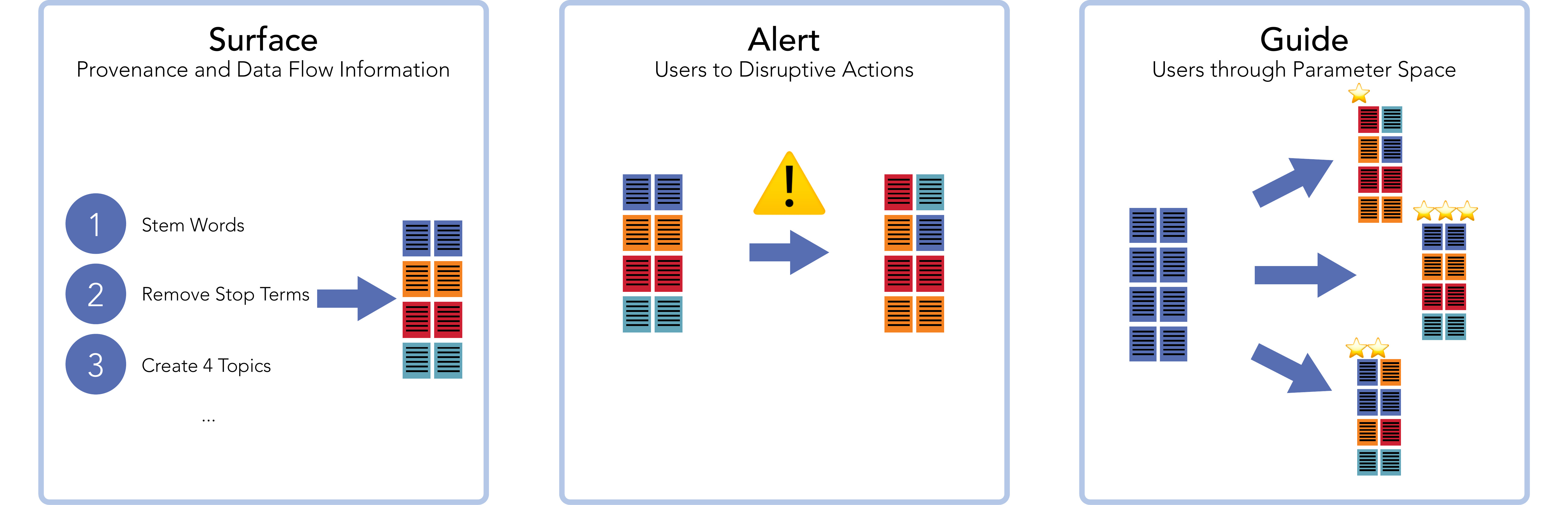}
    \caption{Our suggestions for design, motivated by our simulation results. Pre-processing and data prep steps can be tremendously influential to the resulting model, and so these decisions should be \textbf{surfaced} to the user. What users \textit{think} will be the impact of their choices, and the \textit{actual} scale of impact may be mis-aligned; systems should \textbf{alert} users of these mismatches. Lastly, the parameter space can be large and result in dramatically different results: the system should be proactive and explore some of these analytical paths ahead of time, and \textbf{guide} the user to fruitful areas of the parameter space.}
    \Description{Abstract diagrams of our three categories of design recommendations. They are labeled: 1. Surface provenance and data flow information, 2. alert users to disruptive actions, and 3. guide users through parameter space.}
    \label{fig:suggestions}
\end{figure*}

Our findings above point to three potential implications for designers of future human-in-the-loop topic modelling systems (see \autoref{fig:suggestions}):

The need to \textbf{surface} provenance and data flow information. Given the complexities and degrees of freedom involved in processing text, differing options such as how to tokenize, stem, and filter texts are often not visualized in systems. These actions are either taken via smart defaults from the system, or left as options to power users hidden behind command line interfaces or low level libraries. These preparation-related actions were the most influential in our simulations, and we believe deserve additional consideration when visualizing topic models. Focusing on just user actions after the data preparation state such as cluster manipulation may be just an example of ``rearranging deck chairs on the Titanic:'' a large portion of the descriptive success or failure of the model may have already been decided by earlier, hidden decisions. The lack of provenance visualization has been portrayed as an ethical concern in current visualization practices~\cite{correll2019ethical}. In topic modelling this deficiency is also a practical and pragmatic concern: without knowing how texts were prepared, it is difficult to compare or interpret topic models, especially across different states. 

The need to \textbf{alert} users to the potential impact of their actions on the model. It was not clear to us, \textit{a priori}, which actions would have large impacts on the resulting model (hence our lack of strong stated hypotheses to this effect); our suspicions were guided by point experiences and folklore. We expect that many potential analysts using topic models are in a similar situation. This lack of accurate intuitions, combined with the potential lack of visibility of the effects of actions, can create potential mismatches (algebraic or otherwise) in what the user intends to happen as the result of an action, and what actually results in the topic model. This suggests that designers of systems should employ testing or other regimes to flag potential mismatches, and surface these results to the users. We believe that our metrics and simulation pipeline provide natural support to this sort of user experience: the system could proactively calculate the impact of an action, and report the scale of this impact to the user. At the very least, we would caution designers from providing only one view of the topic model at a time: one individual perspective of a topic model (such as top terms, or top documents) may not suffice to reliably present what has changed or remained the same after a user action.

The need to \textbf{guide} users to help them decide amongst potential actions, or to explore (potentially analytically fruitful) paths not taken. El-Assady et al.~\cite{el2018visual} is an example of what this sort of guidance might look like in a text analytics context: the system proposes \textit{optimizations} based on \textit{speculative execution} of particular branches of the parameter space. The direct exposure of the ``forking paths''~\cite{pu2018garden} or ``multiverse''~\cite{liu2020boba} of analyses could allow users to take ownership of the model while still being cognizant of changes to model structure or performance. A human-in-the-loop system need not simulate the entire complex parameter space, but, as in Lee et al.'s~\cite{lee2019human} ``cruise control'' metaphor, be guided by the user to particular areas, and then perform local exploration of parameter space to find areas with the best outcomes. In such a regime, it is also possible that designers will need to more tightly integrate uncertainty information into their topic model visualizations, as topic and text data could shift in ways that the model might be unable to predict.

\subsection{Limitations \& Future Work}

Our simulations were relatively modest in scope, exploring the impact of only one individual action at a time.  In a fully expressive human-in-the-loop topic modelling system, users would likely undertake a series of actions as they iteratively refine the data model and resulting topics. Simulating these actions requires both a concatenation of actions, and a more purposeful selection of tokens, topics, and texts upon which to operate. These actions are unlikely to be commutative, and thus the simulation of complex chains of user actions presents a combinatorial and analytical challenge. How this challenge is managed depends largely on how and when a pipeline or model is updated in response to a user action. Similarly, while we have deployed our simulation pipeline across multiple datasets, in this work we report mainly on one dataset. Although the $S_{r}$ score we use in our analysis appears able to transfer to other datasets, we already show here that impact of individual user actions is still dataset dependent. We encourage readers to examine the effects of user actions on their own datasets and caution against generalizing our findings to all possible text corpora. We have provided our pipeline as a means to do so and have developed it in such a way to extend to more complex and sequential user actions. %Many of the impacts we describe are likely to be heavily dataset dependent. We would caution readers against generalizing our findings across all possible text corpora. 

We see three immediate open avenues for future exploration. The first is to refine our results in order to construct and validate summary scales or metrics that can capture the many ways in which a topic model can change as a result of user actions. There are many potentially richer metrics to measure the coherency and user surprise engendered by a particular topic model. Richer models of impact would provide us with more confidence in making our proposed interventions (for instance, usefully alerting users to disruptive changes). On the other hand, a single (or a small set) of holistic, well-validated, informative, topic model change metrics would afford more streamlined communication between the user and the system, and provide guidance between alternative model choices even in unsupervised or partially-supervised settings.

Secondly, judging from our visual analyses, the connection between metric impact and human judgments or perception of that impact as instantiated in particular visualization tools is unclear. In future work, we intend on conducting human subjects experiments to anchor our suppositions about ``noticeable'' or ``important'' actions to human judgments, both in terms of our summary metrics but also in terms of our AVD analyses. More rigorous and human-centric analyses of impact could suggest more ``robust'' or ``defensive'' visualizations of topic models, or more proactive or collaborative topic modelling user experiences.

Lastly, we hope that this work points to the potential of simulation work to augment existing practices around more traditional user studies. Within the constraints of a short term user study, participants may only be able to explore a small portion of a total interaction space. Simulating these actions could be used to identify scenarios and settings in need of particular attention from follow-on user studies, or provide reasonable approximations in areas where user data are missing. Simulations could even be used to create models of the users' mental models directly (as per the call in Kumar et al.~\cite{kumar2019didn} to include informed priors in human-in-the-loop topic models), allowing a better channel of communication between human and algorithm.

%ecological validity - impact of datasets\\
%sequencing multiple events (this quickly becomes a combinatorial problem)\\
%Future work : a user study, which would help us assess the user's percieved impact of action and the model. \\
%Future work : a summary metric for the user's action (trickey here, because stuff is dataset dependent too.

\subsection{Conclusion}

In this work we use simulation as a design probe to explore the impact of potential user actions on an abstract human-in-the-loop topic modelling pipeline. We find that user actions are unevenly disruptive to these models, in ways that are not adequately captured by existing topic model visualizations or interactive systems. Our findings are important for designers who wish to leverage human knowledge and agency in their systems. Moreover, we believe that these results point to new and exciting research opportunities to realize the potential of human-in-the-loop text analytics through new metrics, visualization strategies, and user experiences.